\documentclass[journal,comsoc]{IEEEtran}
\ifCLASSOPTIONcompsoc
  \usepackage[nocompress]{cite}
\else
  \usepackage{cite}
\fi
\ifCLASSINFOpdf
\else
\fi
\usepackage{amssymb}
\usepackage{graphicx}
\usepackage{xcolor}
\usepackage{todonotes}
\usepackage{enumitem}
\usepackage{url}
\usepackage{epstopdf}
\usepackage{subfigure}
\usepackage[maxfloats=40]{morefloats}
\usepackage{multirow}
\usepackage{amsmath}
\usepackage{tabularx}
\usepackage{mathtools}
\usepackage{verbatim}
\usepackage{balance}
\usepackage{float}
\usepackage{hhline}
\usepackage{multirow}
\usepackage{array}
\usepackage{pifont}
\usepackage{makecell}
\usepackage[colorlinks=true, allcolors=black]{hyperref}
\newcolumntype{Y}{>{\centering\arraybackslash}X}
\newcolumntype{M}[1]{>{\centering\arraybackslash}m{#1}}
\newcolumntype{?}{!{\vrule width 1pt}}

\newtheorem{definition}{Definition}

\begin{document}

\title{FIVR: Fine-grained Incident Video Retrieval}

\author{Giorgos Kordopatis-Zilos, ~\IEEEmembership{Student Member,~IEEE,}
		Symeon Papadopoulos, ~\IEEEmembership{Member,~IEEE,}
        Ioannis Patras, ~\IEEEmembership{Senior Member,~IEEE,}
        Ioannis (Yiannis) Kompatsiaris, ~\IEEEmembership{Senior Member,~IEEE}
\thanks{G. Kordopatis-Zilos is with the Information Technologies Institute (ITI), Centre for Research and Technology Hellas (CERTH), 6th km Charilaou - Thermi, 57001, Thessaloniki, Greece, and Queen Mary University of London, Mile End road, E1 4NS London, UK. e-mail: georgekordopatis@iti.gr

S. Papadopoulos and Y. Kompatsiaris are with ITI-CERTH, 6th km Charilaou - Thermi, 57001, Thessaloniki, Greece. e-mails: \{papadop, ikom\}@iti.gr

I. Patras is with Queen Mary University of London, Mile End road, E1 4NS London, UK. e-mail: i.patras@qmul.ac.uk

This paper has supplementary downloadable material available at http://ieeexplore.ieee.org., provided by the author. The material includes [brief description on what the material includes and shows]. This material is XXXX in size.
}}
\maketitle

\begin{abstract}
This paper introduces the problem of Fine-grained Incident Video Retrieval (FIVR). Given a query video, the objective is to retrieve all associated videos, considering several types of associations that range from duplicate videos to videos from the same incident. FIVR offers a single framework that contains several retrieval tasks as special cases. To address the benchmarking needs of all such tasks, we construct and present a large-scale annotated video dataset, which we call FIVR-200K, and it comprises 225,960 videos. To create the dataset, we devise a process for the collection of YouTube videos based on major news events from recent years crawled from Wikipedia and deploy a retrieval pipeline for the automatic selection of query videos based on their estimated suitability as benchmarks. We also devise a protocol for the annotation of the dataset with respect to the four types of video associations defined by FIVR. Finally, we report the results of an experimental study on the dataset comparing five state-of-the-art methods developed based on a variety of visual descriptors, highlighting the challenges of the current problem.

\end{abstract}


\section{Introduction}
\label{sec:introduction}

Video retrieval is a very important, yet highly challenging problem that is exacerbated by the massive growth of social media applications and video sharing platforms. As a result of the uncontrolled number of videos published in platforms, such as YouTube, it is very common to find 
multiple videos about the same incident (e.g., terrorist attack, plane crash), 
which are either near-duplicates of some original video, or simply depict the same incident from different viewpoints or at different times. Being able to efficiently retrieve all videos around an incident of interest is indispensable 
for numerous applications ranging from copy detection for copyright protection\cite{douze2010,kraaij2011} to event reconstruction\cite{browne2017,mairs2018,gao2017} and news verification\cite{jin2017,silverman2013}.

However, different instances of the video retrieval problem pose different requirements.
In the copy detection problem, 
given a query video, 
only videos containing nearly identical copies of the video should be retrieved. In such a scenario, similar videos from the same incident should be considered irrelevant. 
However, tasks such as 
journalistic investigations around an
incident pose different requirements.  Being able to efficiently and accurately retrieve i) videos that originate from the same video source (duplicate videos), and 
ii) videos that capture the same incident from different viewpoints and at different times, would be of great value for such tasks. 
In this paper, we denote the overall 
problem as Fine-grained Incident Video Retrieval (FIVR) and construct a large scale dataset to simulate its different instances. 

\begin{figure*}[t]
\centering
\includegraphics[width=17.4cm]{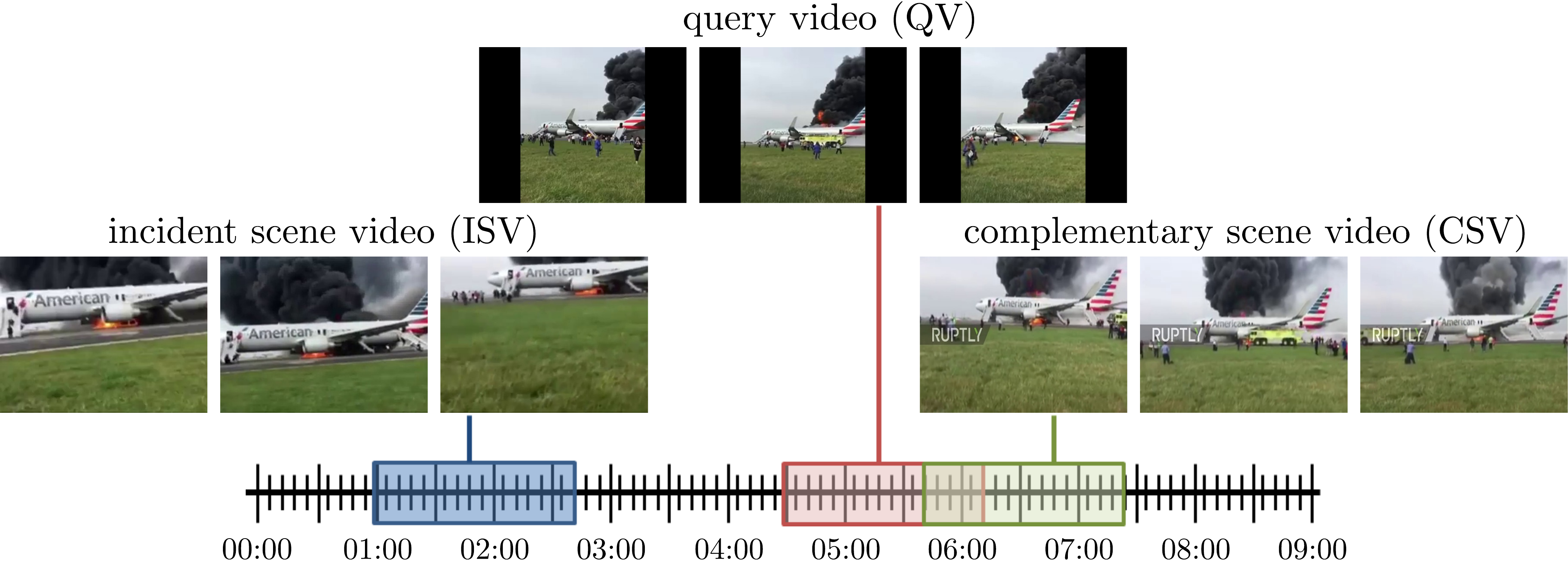}
\caption{Examples of a query video (QV) with one \textit{complementary scene video} (CSV) and one \textit{incident scene video} (ISV) on the timeline of an incident. The following colour coding is used: i) red for QV, ii) green for CSV, and ii) blue for ISV.}
\label{fig:timeline}
\end{figure*}

There are several application areas where the FIVR problem 
can prove relevant. 
A number of such relevant retrieval applications are presented in \cite{feng2013}. For instance, news media analysis and reporting 
would greatly benefit from an effective solution to the FIVR problem. In a recent work, journalists from the \textit{New York Times} \cite{browne2017} managed to reconstruct the Las Vegas shootings based on content from both amateur and police videos that had been captured during the incident. In another relevant work, the research group \textit{Forensic Architecture} \cite{mairs2018} created a 3D video of the Grenfell Tower fire to help understand how the disaster unfolded. Moreover, Gao et al. \cite{gao2017} developed an approach that automatically processes a set of collected web videos and generates a short video that summarizes the storyline of an event. 
Other application scenarios and use cases that may benefit from solutions to the FIVR problem include safety and security applications \cite{sabeur2015,popoola2012,lipton2004}. 
Such applications could considerably benefit from methods that, given a query video, retrieve similar videos, based on the different definitions of FIVR association.

We address two fundamental associations between similar videos: a) duplicate videos, and b) videos of the same incident. By duplicate videos we refer to videos that have been captured by the same camera and depict exactly the same scene, but may have undergone some visual transformations (e.g., brightness/contrast, colour, recompression, noise addition, cropping). 
The second type of similar videos that we consider are videos capturing the same incident. This category may be split into subcategories: a) videos that depict the same incident scene from complementary viewpoints, and b) videos that capture the same incident at different time intervals. In particular, two videos in the first category must have at least one video segment where there is temporal overlap between the depicted incident. 
Videos in the second subcategory need to depict the same incident but do not need to have temporal overlap. Figure \ref{fig:timeline} illustrates 
three example videos that capture the same incident along with their FIVR associations.

The goal of this paper is to propose and formulate the Fine-grained Incident Video Retrieval (FIVR) problem through the composition of a challenging dataset that will serve the benchmarking needs for different variants of the problem. To accurately represent the problem, this dataset is composed of user-generated videos related to a large number of real-world events. The events were selected to be of the same nature for the collected videos to be visually similar and thus to include more challenging distractors in the dataset. Moreover, 
a number of videos have been selected as queries through a principled process. The ideal benchmark query should have many duplicates and videos from the same incident, but at the same time, 
there should also be many visually similar distractor videos from different events to make the retrieval of relevant videos more challenging.

The main contributions of this work can be summarized in the following:
\begin{itemize}
\item The introduction of the Fine-grained Incident Video Retrieval (FIVR) problem and the definition of different associations between pairs of videos. 
\item The creation and availability of a large-scale dataset (FIVR-200K)\footnote{\url{http://ndd.iti.gr/fivr/}} consisting of 225,960 videos.
\item The development of a process for the collection and annotation of videos based on major news events crawled from Wikipedia and a principled process for the automatic selection of suitable video queries. 
\item A comprehensive experimental study comparing five state-of-the-art approaches implemented with several visual descriptors (handcrafted and deep features).
\end{itemize}
The rest of the paper is organized as follows. Section \ref{sec:soa_datasets} presents related research and datasets. Section \ref{sec:definitions} introduces the necessary notation and definitions. Section \ref{sec:dataset} describes the dataset construction process, including the video collection, query selection and result annotation. Section \ref{sec:benchmark} reports on the results of the experimental study on the dataset. Section \ref{sec:conclusion} concludes the paper.

\section{Related Work}
\label{sec:related_work}

\subsection{Video Datasets}
\label{sec:soa_datasets}

There is a variety of retrieval tasks and definitions in the multimedia community in relation to the FIVR problem. These vary with respect to the degree of similarity that determines whether a pair of videos are considered related, and range from Near-Duplicate Video Retrieval (NDVR) with very narrow scope where only almost identical videos are considered positive pairs \cite{wu2007}, to very broad definitions, where videos from the same event \cite{revaud2013} 
or with the same semantics \cite{basharat2008} are labelled as related. However, there does not seem to be strong consensus among researchers about which videos are considered near-duplicate videos and none of the existing definitions addresses the retrieval of \textit{same incident videos}; 
in this paper, we attempt to address these issues and provide solid definitions for all types of associations between videos related to the FIVR problem.

Additionally, although there are a few video collections that capture different aspects of this problem, all of them are limited in different ways. More specifically, related datasets include CC\_WEB\_VIDEO \cite{wu2007}, UQ\_VIDEO \cite{song2011}, MUSCLE-VCD \cite{law2007}, TRECVID-CBCD 2011 \cite{kraaij2011}, VCDB \cite{jiang2014} and EVVE \cite{revaud2013}. The first two datasets were collected for the problem of near-duplicate video retrieval, the next three for the video copy detection problem, and the last one for the problem of event retrieval. The query videos for the MUSCLE-VCD and TRECVID-CBCD datasets were artificially generated, whereas the rest of the datasets contain actual user-generated videos as queries. Table \ref{tab:datasets} provides an overview of the aforementioned datasets and associated retrieval tasks.

%


\begin{table*}[t]
\centering
\small
\caption{Comparison of FIVR with existing datasets and retrieval tasks.}
\label{tab:datasets}
\begin{tabular}{|l|c|c|c|c|l|}
\hline 
\textbf{Dataset} & \textbf{Queries} & \textbf{Videos} & \textbf{Hours} & \textbf{User-generated} & \textbf{Retrieval Task} \\ \hhline{======}
CC\_WEB\_VIDEO   \cite{wu2007}      & 24     & 12,790   & 551   & \checkmark & Near-Duplicate Video Retrieval \\ \hline
UQ\_VIDEO        \cite{song2011}    & 24     & 169,952  & N/A   & \checkmark & Near-Duplicate Video Retrieval \\ \hline
MUSCLE-VCD       \cite{law2007}     & 18     & 101      & 100   & \ding{55}  & Video Copy Detection           \\ \hline
TRECVID 2011     \cite{kraaij2011}  & 11,256 & 11,503   & 420   & \ding{55}  & Video Copy Detection           \\ \hline
VCDB             \cite{jia2014}     & 528    & 100,528  & 2,038 & \checkmark & Partial Video Copy Detection   \\ \hline
EVVE             \cite{revaud2013}  & 620    & 102,375  & 5,536 & \checkmark & Event Video Retrieval          \\ \hhline{======}
FIVR-200K                           & 100    & 225,960  & 7,100 & \checkmark & Fine-grained Incident Video Retrieval   \\ \hline
\end{tabular}
\end{table*}

The most relevant, publicly available and widely used dataset is the CC\_WEB\_VIDEO \cite{wu2007}. The dataset consists of user-generated videos collected from the Internet; in particular, it contains a total of 12,790 videos consisting of 397,965 keyframes. The videos were collected by submitting 24 popular text queries to popular video sharing websites (YouTube, Google Video, and Yahoo! Video). For every query, a set of video clips were aggregated, and the most popular video was considered as the query video. 
Subsequently, all retrieved videos in the video sets were manually annotated by three annotators based on their near-duplicate relation to the query video. 
The near-duplicate rate of the collected sets ranges from 6\% to 93\%. On average,  27\% of the videos in each set are considered near-duplicates.

Several variations of the CC\_WEB\_VIDEO dataset have been developed by researchers in the NDVR field \cite{shang2010,song2011,cai2011,chou2015}. To make the NDVR problem more challenging and benchmark the scalability of their approaches, researchers usually extend the core CC\_WEB\_VIDEO dataset with thousands of distractor videos \cite{song2011,chou2015}. 
The most well-known and publicly available dataset that has been created through this process is UQ\_VIDEO \cite{song2011}. For the composition of the background dataset, they chose the 400 most popular queries based on Google Zeitgeist Archives from the years 2004 to 2009. Each query was submitted to YouTube and up to 1,000 video results were collected. After filtering out videos of duration longer than 10 minutes, the combined dataset is composed of 169,952 videos (including those of the CC\_WEB\_VIDEO) comprising 3,305,525 keyframes. The same 24 query videos contained in CC\_WEB\_VIDEO are used for benchmarking. 

Another popular public dataset is the MUSCLE-VCD, created by Law-To et al. \cite{law2007}. This dataset was created for the problem of video copy detection. It consists of 100 hours of videos including Web video clips, TV archives, and movies of different bitrates, resolutions and video formats. A set of original videos and their corresponding transformed queries are given for evaluation. Two types of transformation are applied on the queries: a) ST1: copy of the entire video with a single transformation, where the videos may be slightly recoded and/or subjected to noise addition; b) ST2: partial copy of videos, where two videos share one or more video segments. Both transformations were artificially applied using video-editing software. The transformed videos or segments were used as queries to search their original versions in the dataset.

The annual TRECVID \cite{trecvid} evaluation included a task on copy detection in years 2008 to 2011. Each year a benchmark dataset was generated and released only to the registered participants of the task. The TRECVID datasets were constructed following the same process as the MUSCLE-VCD dataset. The latest edition of the dataset \cite{kraaij2011} contains 11,503 reference videos of over 420 hours and 11,256 queries. Query videos are categorized into three types: a reference video only, a reference video embedded into a non-reference video, and a non-reference video only. Only the first two types of query video are copies of videos in the dataset. The queries were automatically generated by randomly extracting a segment from a dataset video and imposing a few predefined transformations. The contestants were asked to find the original videos and detect the copied segment.

A more recent dataset that is relevant to our problem is VCDB \cite{jia2014}. It is composed of videos from popular video platforms (YouTube and Metacafe) and has been compiled and annotated as a benchmark for the partial copy detection problem. VCDB contains two subsets, the \textit{core} and \textit{distractor}. The core subset contains 28 discrete sets of videos composed of 528 videos with over 9,000 pairs of partial copies. Each video set was manually annotated by seven annotators, and the video chunks of the video copies were extracted. The distractor subset is a corpus of approximately 100,000 distractor videos that is used to make the video copy detection problem more challenging. In total, VCDB contains 100,528 videos amounting to more than 2,000 hours of video.

Finally, the EVVE 
dataset \cite{revaud2013} was developed for the problem of event video retrieval. The main task 
on this dataset is the retrieval of all videos that capture the event depicted by a query video. The dataset contains 13 major events that were provided as queries to YouTube. A total of 2,375 videos were collected, and 620 of them were selected as queries. Each event was annotated by one annotator, who first produced a precise definition of the event. In addition to the videos collected for the specific events, the authors also retrieved a set of 100,000 distractor videos by querying YouTube with unrelated terms. These videos were all collected before a certain date, which ensures that the distractor set does not contain any of the relevant events of EVVE since all events occurred after that date.

All aforementioned datasets have limitations. 
For instance, the volume and query set of CC\_WEB\_VIDEO are relatively small (12,790 videos and 24 queries), and the dataset lacks challenging distractors given that the queries are very 
different from each other. 
The main limitation of TRECVID-CBCD 2011 
is that the video copies are artificially generated by applying standard transformations to a corpus of videos. Regarding the VCDB dataset, 
only a limited number of its videos have been annotated (528 videos in the core dataset). In EVVE, the definition of the related videos is much more fuzzy, 
and additionally, the dataset contains annotations only for videos from the same event, and not for its near-duplicates. 
In short, none of the above datasets can satisfy the requirements posed by the FIVR problem. For that reason, we built a new large-scale video dataset (FIVR-200K) according to the FIVR definition. The dataset consists of videos depicting a variety of real-world news events, challenging cases of positive video pairs, and a large number of distractor videos. 

\subsection{Video Retrieval Methods}
\label{sec:soa_methods}

There are several related works in the literature dealing with the problem of similarity-based video retrieval. In general, a typical video retrieval framework consists of two key components: (i) feature extraction, where visual descriptors are extracted from video frames, (ii) feature aggregation and similarity calculation, where frame descriptors are processed to calculate the video similarity between the query and all videos in the dataset. In this context, we present the main trends in the literature for each component separately.

There is a wide variety of features that have been crafted to capture the visual information of video content. A common strategy is to extract some frames from videos via uniform sampling and then extract their visual descriptors based on the global and/or local information contained in the frames. Early approaches employed  handcrafted features including HSV Colour Histograms \cite{wu2007,song2011,hao2017,jing2018}, Local Binary Patterns (LBP)\cite{zhao2007} \cite{shang2010,song2011,wu2014,jing2018}, fuzzy multidimensional histograms of colour and motion video segments \cite{doulamis2000}, Auto Colour Correlograms (ACC) \cite{huang1997, cai2011}, and several keypoint descriptors (such as SIFT \cite{lowe2004,wu2007,zhao2009,jiang2007} and SURF \cite{bay2006, chou2015}) combined with Vector of Locally Aggregated Descriptors (VLAD) \cite{jegou2010,revaud2013}. Additionally, several recent approaches in related video retrieval fields have employed features extracted from the activations of deep Convolutional Neural Network (CNN) architectures due to their high effectiveness. Common feature extraction techniques include the extraction from the activations of one of the fully connected layers of CNN architectures \cite{jiang2016,wu2018}, the application of the Maximum Activation of Convolution (MAC) pooling function on the activations of the intermediate CNN layers \cite{kordopatis2017a,kordopatis2017b}, or using Regional Maximum Activations of Convolutions (RMAC) \cite{tolias2016} based on the output of a Region Of Interest (ROI) pooling layer applied on the final convolutional layer \cite{garcia2018, baraldi2018}.

A wide variety of feature aggregation and similarity calculation schemes have been implemented. One of the earliest schemes is the generation of a global vector, where all frame descriptors are averaged to a single vector for the entire video. Video similarity is then calculated based on the dot product between the respective vectors, as proposed in \cite{wu2007}. A very popular feature aggregation technique among the research community is the Bag-of-Words (BoW) scheme. Every frame descriptor is mapped to one or more visual words and the final video representation is the \textit{tf-idf} representation of these visual words. Video ranking is performed based on the cosine similarity between the tf-idf representations. Variants of the BoW scheme have been used in \cite{jiang2007,chou2015,cai2011,kordopatis2017a,zhao2009,shang2010}. For instance, Layer Bag-of-Words (LBoW) \cite{kordopatis2017a} is a variant of the BoW scheme based on the intermediate CNN features, where the feature vectors extracted from each convolutional layer are mapped to a  word of a visual codebook (a different codebook is generated per layer), and then aggregated using tf-idf. 
Another popular aggregation practice is the generation of a hash code that represents the entire video. Such methods usually combine multiple image features to learn a group of hash functions that project video frames into the Hamming space and then combine them to a single video representation. Hamming distance is employed to determine the similarity between videos. Some representative approaches include \cite{song2011,hao2017,jing2018,song2013,song2018,wu2018}. Finally, there are some recent works that 
employ supervised learning for improved
video similarity calculation, most often relying on Deep Metric Learning (DML). 
A network is fed with pairs or triplets of videos and is trained based on a loss function that minimizes the distance between related videos and maximizes the distance between irrelevant videos. Jiang et al. \cite{jiang2016} trained a CNN for partial video copy detection using the pairwise contrastive loss function that minimizes the distance between sampled frame patches and their manually transformed versions and maximizes the distance between irrelevant frame patches. They employ Euclidean distance to measure frame distances and a temporal alignment method to detect copied video segments. In \cite{kordopatis2017b}, a deep neural network is trained 
to find an embedding function that maps videos in a feature space where near-duplicates are closer to each other than to irrelevant videos. The distance between videos is determined by their Euclidean distance in the embedding space. In another work, Baraldi et al. \cite{baraldi2018} introduced a temporal layer in a deep network that calculates the temporal alignment between videos. They trained the network minimizing the triplet loss that takes into account both localization accuracy and recognition rate.

\section{Definitions}
\label{sec:definitions}

\begin{table*}[t]
\centering
\normalsize
\caption{Definitions of the different types of associations between video pairs.}
\label{tab:definitions}
\small
\begin{tabular}{|m{2cm}|m{3cm}|m{11.5cm}|}
\hline
\textbf{Duplicate Scene Videos (DSV)} 
&
Videos that share at least one scene (captured by the same camera) regardless of any applied transformation.
&
\definition 
\label{def:dsv}
Given a query video $\textbf{q}$ with a number of $n$ scenes $\textbf{q}=[q_1\ q_2\ ...\ q_n]$, spatio-temporal span $\textbf{z}^q$ and viewpoints $\textbf{v}^q$, and a candidate video $p$ with a number of $m$ scenes $\textbf{p}=[p_1\ p_2\ ...\ p_m]$, spatio-temporal span $\textbf{z}^q$ and viewpoints $\textbf{v}^p$, there is a binary function DS$(\cdot, \cdot)$ that indicates whether the two videos are DSVs
\begin{equation}
\text{DS}(q,p)=
  \begin{cases}
    1   &  \exists i \in [1,m]\ (z^p_{i} \subseteq \textbf{z}^q \land v^p_{i} \in \textbf{v}^q)\\
    0   &  \text{otherwise}
  \end{cases}
\label{eq:dsv}
\end{equation} \\ \hline
\textbf{Complementary Scene Videos (CSV)}
&
Videos that contain part of the same spatio-temporal segment, 
but captured from different viewpoints.
&
\definition 
\label{def:csv}
Given a query video $\textbf{q}$ with a number of $n$ scenes $\textbf{q}=[q_1\ q_2\ ...\ q_n]$, spatio-temporal span $\textbf{z}^q$ and viewpoints $\textbf{v}^q$, and a candidate video $p$ with a number of $m$ scenes $\textbf{p}=[p_1\ p_2\ ...\ p_m]$, spatio-temporal span $\textbf{z}^q$ and viewpoints $\textbf{v}^p$, there is a binary function CS$(\cdot, \cdot)$ that indicates whether the two videos are CSVs.
\begin{equation}
\text{CS}(q,p)=
  \begin{cases}
    1   &  \exists i \in [1,m] \ (z^p_{i} \subseteq \textbf{z}^q \land v^p_{i} \notin \textbf{v}^q)\\
    0  &  \text{otherwise}
  \end{cases}
\label{eq:csv}
\end{equation} \\ \hline
\textbf{Incident Scene Videos (ISV)}
&
Videos that capture the same incident, i.e. they are spatially and temporally close, but have no overlap.
& 
\definition 
\label{def:isv}
Given a query video $\textbf{q}$ with a number of $n$ scenes $\textbf{q}=[q_1\ q_2\ ...\ q_n]$, spatio-temporal span $\textbf{z}^q$ and incidents $\textbf{h}^q$, and a candidate video $p$ with a number of $m$ scenes $\textbf{p}=[p_1\ p_2\ ...\ p_m]$, spatio-temporal span $\textbf{z}^q$ and incidents $\textbf{h}^p$, there is a binary function IS$(\cdot, \cdot)$ that indicates whether the two videos are ISVs.
\begin{equation}
\text{IS}(q,p)=
  \begin{cases}
    1   &  \exists i \in [1,m] \ h^p_{i} \in \textbf{h}^q \land \nexists j \in [1,n] \ z^p_{j} \subseteq \textbf{z}^q\\
    0  &  \text{otherwise}
  \end{cases}
\label{eq:ndv}
\end{equation} \\
\hline
\end{tabular}
\end{table*}

\begin{table}[h]
  \centering
  \normalsize
  \caption{Background notation and definitions.
  }
  \label{tab:glossary}
  \begin{tabular}{|M{0.8cm}|m{6cm}|}
    \hline
    \textbf{Term}   & \textbf{Description} \\ \hhline{==}
    $\textbf{x}$    & an arbitrary video \\ \hline
    $x_i$           & $i^{\textnormal{th}}$ scene of $\textbf{x}$ \\ \hline
    $z_i^x$         & spatio-temporal span of the $i^{\textnormal{th}}$ scene of $\textbf{x}$ \\ \hline
    $v_i^x$         & viewpoint of the $i^{\textnormal{th}}$ scene of $\textbf{x}$ \\ \hline
    $h_i^x$         & incident captured in the $i^{\textnormal{th}}$ scene of $\textbf{x}$ \\ \hline
    $\textbf{z}^x$         & spatio-temporal span of the entire video $\textbf{x}$ \\ \hline
    $\textbf{v}^x$         & viewpoints of the entire video $\textbf{x}$ \\ \hline
    $\textbf{h}^x$         & incidents captured in the entire video $\textbf{x}$ \\ \hline
    $S$             & space of scenes \\ \hline
    $Z$             & space of spatio-temporal span \\ \hline
    $V$             & space of viewpoint \\ \hline
    $H$             & space of incidents \\ \hline
    $f$             & function that maps an incident to a unique spatio-temporal span \\ \hline
    $g$             & function that, given a viewpoint, maps a spatio-temporal span to a scene \\ \hline
\end{tabular}
\end{table}

We consider that a real-world incident determines a unique spatio-temporal span, i.e. there is a function $f:H \rightarrow Z$ that maps the incidents from an incident space $H$ to a continuous spatio-temporal space $Z$. Furthermore, a video can be perceived as the mapping of the real world to a sequence of two-dimensional raster images with three colour channels. Additionally, as defined in the field of temporal video segmentation \cite{hanjalic1999}, a video can be decomposed in a sequence of \textit{scenes} or \textit{temporal segments}, each covering either a single event or several related events taking place in parallel. Thus, an arbitrary video $\textbf{x}$ with a sequence of $n$ non-overlapping scenes may be denoted as $\textbf{x}=[x_1\  x_2\  ...\  x_n]$, where $x_i \in S$ and $S$ is the space of scenes. 
We may also consider a function $g:Z, V \rightarrow S$ that maps a real-world spatio-temporal span from space $Z$ and given a specific viewpoint from space $V$, where $V$ is the viewpoint space, to a video scene. 
Note that knowing functions $f$ and $g$ is not our objective; instead, they are solely used for the proper formulation of our problem.

For the accurate definition of the associations between videos, we consider that each scene $x_i$ of an arbitrary video $\textbf{x}$ has the corresponding attributes: the captured spatio-temporal span $z^x_{i} \in Z$, the viewpoint $v^x_{i} \in V$ of the camera and the incident $h^x_i\in H$ that corresponds to the captured spatio-temporal span. By aggregating all attributes of the scenes of video $\textbf{x}$, we can derive the attributes for the entire video: the entire captured spatio-temporal span $\textbf{z}^x \in Z$, all viewpoints $\textbf{v}^x \in V$ of the video scenes and the different incidents $\textbf{h}^x \in H$ occurring during the captured spatio-temporal span.


To properly define the relations between videos, we define three fundamental types of association between videos, which are summarized in Table \ref{tab:definitions}. 
These are defined based on the relation between the viewpoints and spatio-temporal spans of the compared videos. 

We denote as \textbf{Duplicate Scene Videos} (DSVs), two videos that share at least one scene (as captured by the same camera) regardless of any applied transformation. The shared scenes must be close to exact duplicates of each other but can be different in terms of photometric variations, editing operations, length, and other modifications. More precisely, they have to originate from the same spatio-temporal span and viewpoint. Videos that contain semantically similar scenes are not considered DSVs. Definition \ref{def:dsv} provides a formal definition of the DSVs. 
A special case of the Definition \ref{def:dsv} is when Equation \ref{eq:dsv} is valid for all scenes of the candidate video. Such cases are denoted as \textbf{Near-Duplicate Videos} (NDVs).

Videos in the second category have to share at least one common segment of the same incident. These are denoted as \textbf{Complementary Scene Videos} (CSVs). In particular, each of the two videos of a CSV pair needs to contain a spatio-temporal segment that is temporally overlapping with the spatio-temporal segment of the other. However, to be included in this category, the two video segments need to be captured from different cameras, and hence, offer complementary viewpoints of the incident. Since the identification of temporal overlap is a challenging task, any audio or visual cue may be taken into consideration to make such an inference. The formal definition of CSVs is provided in Definition \ref{def:csv}.

\begin{figure*}[t]
\centering
\begin{tabular}{cc}
\begin{tabular}{c}
\textbf{Query Video} \\ 
\includegraphics[height=2.2cm]{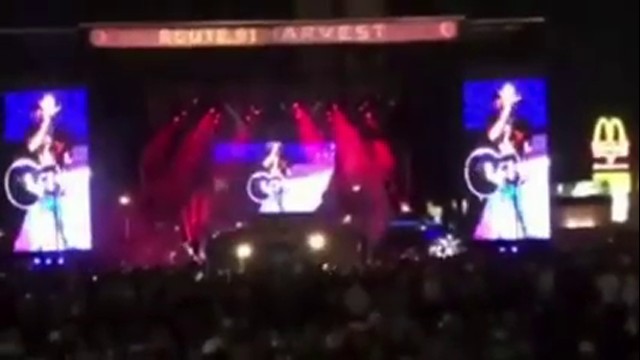} \\
\includegraphics[height=2.2cm]{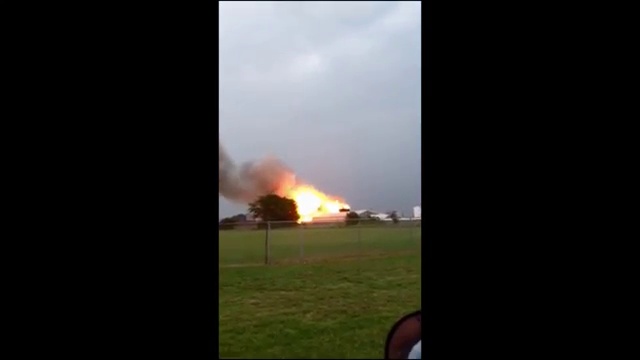} \\
\includegraphics[height=2.2cm]{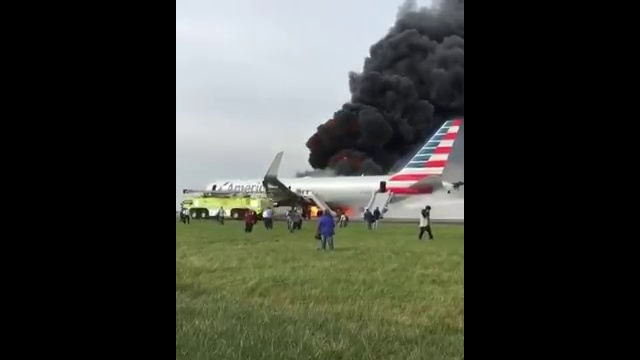} \\
\includegraphics[height=2.2cm]{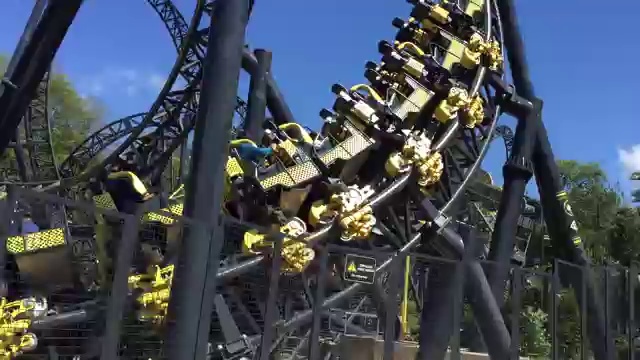} \\
\includegraphics[height=2.2cm]{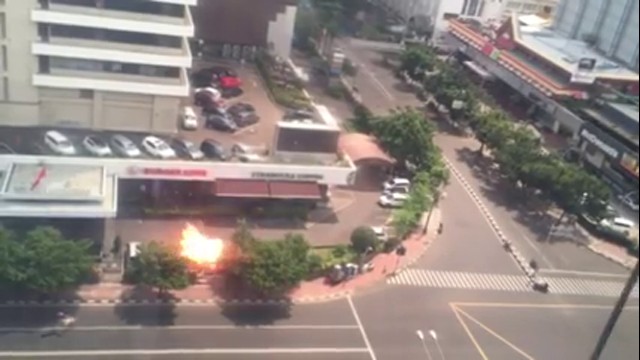} \\ 
\end{tabular}
\begin{tabular}{ccc}
\textbf{Duplicate Scene} & \textbf{Complementary Scene} & \textbf{Incident Scene} \\ 
\includegraphics[height=2.2cm]{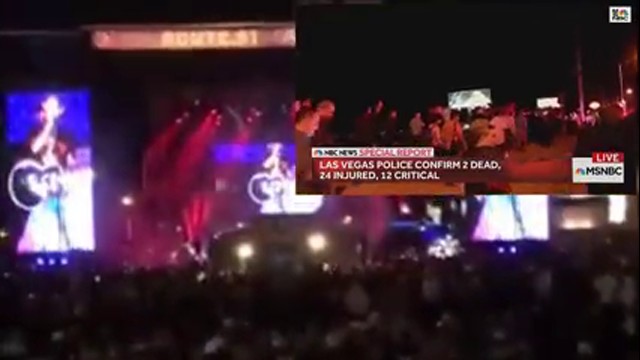} &
\includegraphics[height=2.2cm]{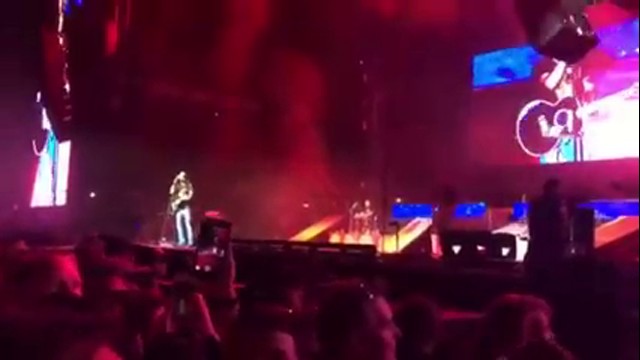} &
\includegraphics[height=2.2cm]{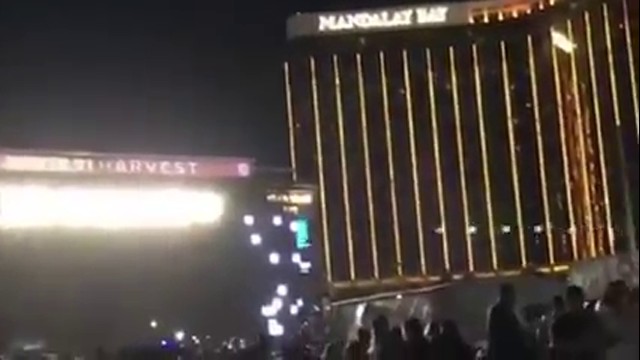} \\
\includegraphics[height=2.2cm]{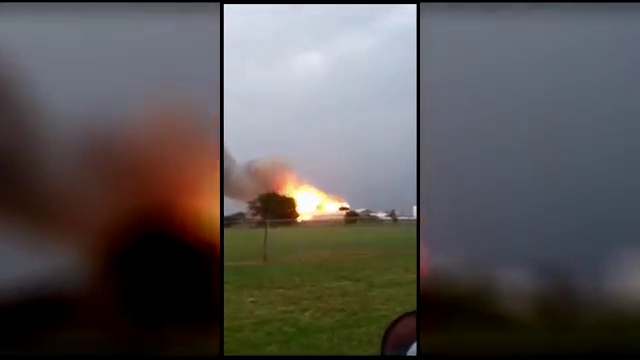} &
\includegraphics[height=2.2cm]{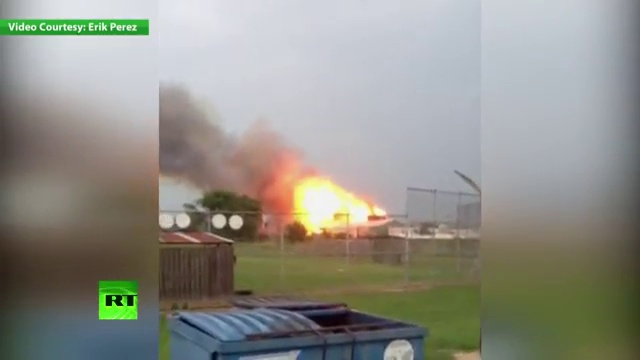} &
\includegraphics[height=2.2cm]{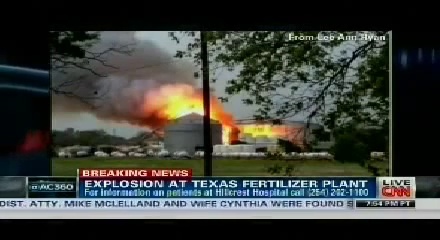} \\
\includegraphics[height=2.2cm]{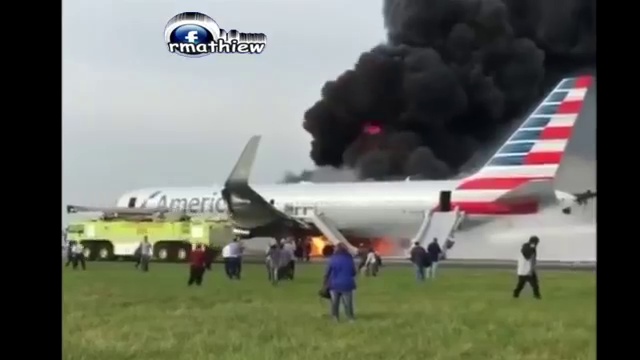} &
\includegraphics[height=2.2cm]{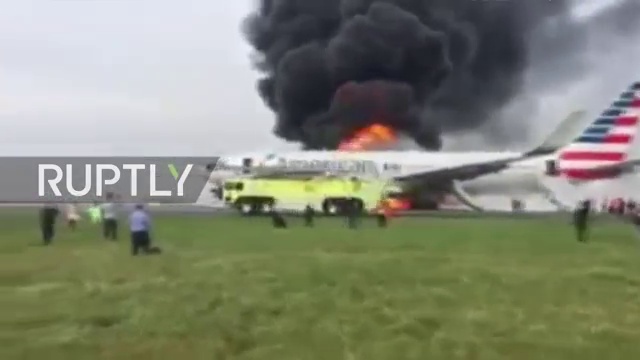} &
\includegraphics[height=2.2cm]{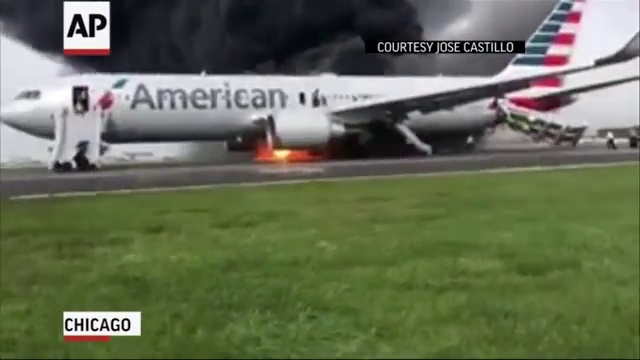} \\
\includegraphics[height=2.2cm]{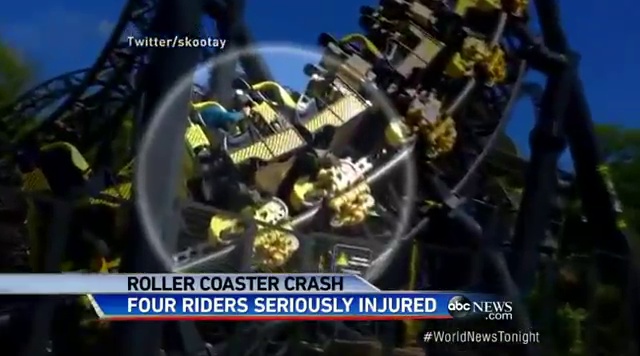} &
\includegraphics[height=2.2cm]{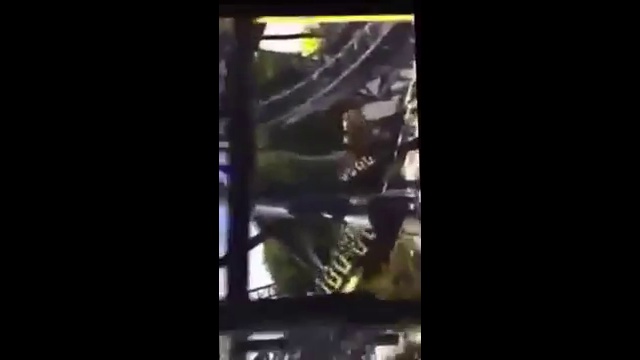} &
\includegraphics[height=2.2cm]{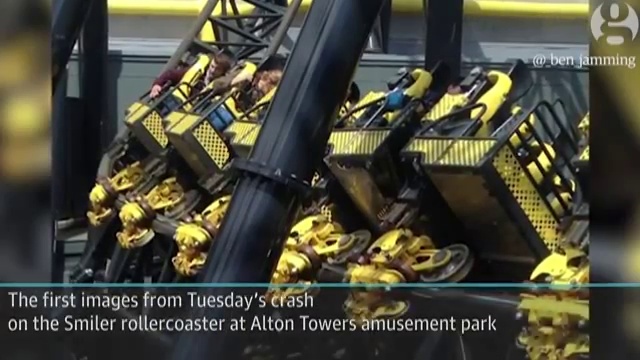} \\
\includegraphics[height=2.2cm]{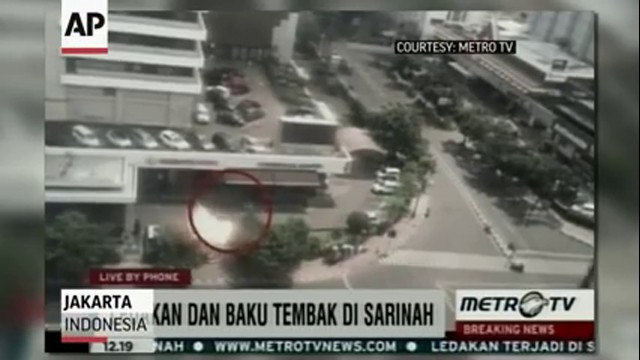} &
\includegraphics[height=2.2cm]{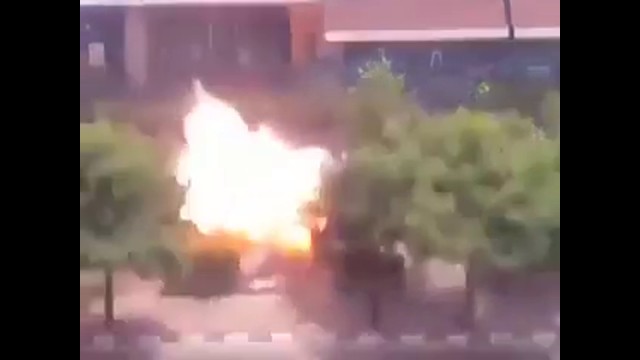} &
\includegraphics[height=2.2cm]{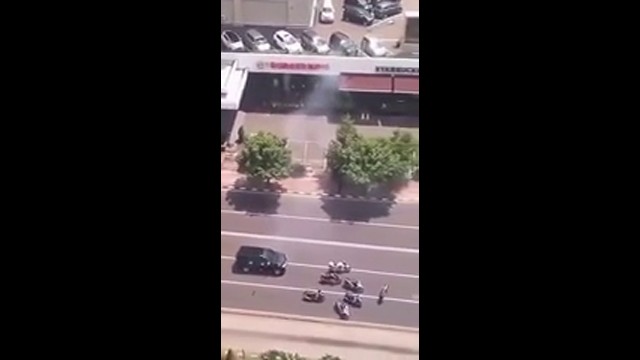} \\ 
\end{tabular}
\end{tabular}
\caption{Examples of queries and retrieved associated videos from FIVR-200K.}
\label{fig:examples}
\end{figure*}

Videos in the third category depict the same incident but have no temporal overlap. These are referred to as \textbf{Incident Scene Videos} (ISVs), and they are formalized 
in Definition \ref{def:isv}. 
Such videos still need to be spatially and temporally related, i.e., their spatio-temporal span should originate from the same incident. Additionally, if the query depicts a particular incident in a long event or a sequence of incidents, then only the videos that capture the particular incident are included in this category. 
Additionally, the inference that two videos originate from the same incident may derive from video metadata (e.g., title, description) or audio, i.e., it is not necessary to associate the two videos with the event solely on the basis of their visual content. There are some rare cases where ISVs have no obvious visual cues linking them to each other, and no such inference can be made without outside knowledge. An example is a case where the query captures an incident from the outside of a building, and there are ISVs from the inside of the same building captured during the same incident. 

Figure \ref{fig:timeline} illustrates selected frames of a query video and one candidate video from each category (CSV, ISV). The video fragments have been coloured accordingly, with red indicating the query video, green the CSV and blue the ISV. Also, a sample timeline is presented to illustrate the time span where each type of video occurs. 
The example video depicts the fire in the American Airlines flight 383 at Chicago O'Hare airport\footnote{\url{https://en.wikipedia.org/wiki/American_Airlines_Flight_383_(2016)}}. There are a number of videos in FIVR-200K from that incident, capturing various viewpoints and different time spans. The query video depicts the passengers standing outside the plane and the firefighters trying to put out the fire. The CSV is captured from a slightly different viewpoint. The overlap between the two videos can be determined from the movement of the firefighter truck passing in front of the plane and the position of the people. The ISV is in a distinct time span relative to the query. It is captured before the query video and at the moment when the passengers exit the plane through the emergency exits. Figure \ref{fig:examples} illustrates some additional examples of FIVR associations. 

\begin{figure*}[t]
\centering
\includegraphics[width=15.7cm]{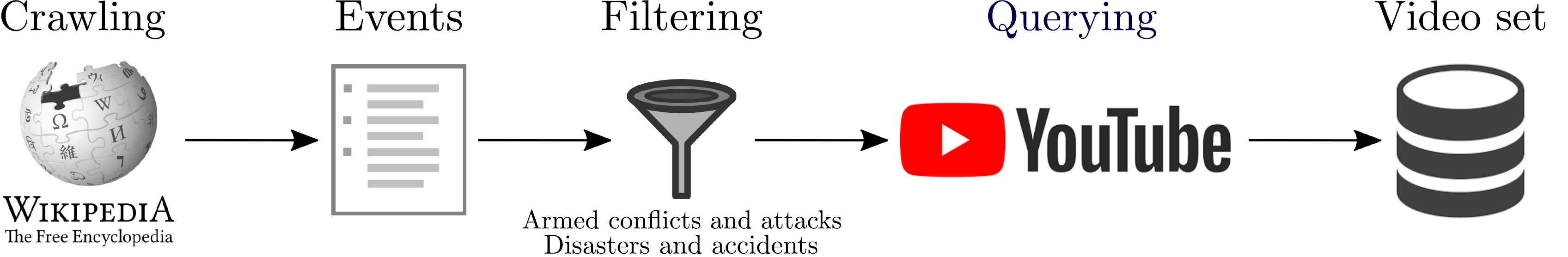}
\caption{Overview of the video collection process.}
\label{fig:overview}
\end{figure*}

\begin{table*}[tbh]
\centering
\caption {Examples of crawled news events.}
\begin{tabular}{lllll}
\textbf{Headline} & \textbf{Date} & \textbf{Category} & \textbf{Text} & \textbf{Source} \\ \hline
Syrian civil war & 2013-01-01 & Armed conflicts and attacks & Fierce clashes erupt near the Aleppo ... & \href{http://www.bbc.co.uk/news/world-middle-east-20882790}{BBC} \\
Greek debt crisis & 2015-07-07 & Business and economics & Eurozone leaders hold a crisis meeting ... &  \href{https://www.reuters.com/article/2015/07/07/us-eurozone-greece-idUSKBN0P40EO20150707}{Reuters} \\
Hurricane Harvey & 2017-08-29 & Disasters and accidents & The death toll from Hurricane Harvey ... & \href{https://www.nytimes.com/2017/08/29/us/hurricane-harvey-storm-flooding.html}{New York Times} \\
United States elections & 2016-11-08 & Politics & Voters in the United States go to polls ... &  \href{http://www.abc.net.au/news/2016-11-08/us-election-what-time-will-we-know-the-result/7943894}{ABC} \\
Artificial intelligence & 2016-01-27 & Science and technology & A computer program called AlphaGo ... & \href{https://www.technologyreview.com/s/546066/googles-ai-masters-the-game-of-go-a-decade-earlier-than-expected/}{MIT Technology Review} \\
Boston Marathon Bombing & 2014-07-21 & Law and Crime & Azamat Tazhayakov, a friend of accused ... & \href{http://news.msn.com/crime-justice/friend-convicted-of-impeding-boston-marathon-probe}{MSN News} \\
2016 Summer Olympics & 2016-08-12 & Sports & Singaporean swimmer Joseph Schooling ... & \href{https://www.nytimes.com/2016/08/13/sports/olympics/michael-phelps-joseph-schooling-200-butterfly-gold.html?_r=0}{New York Times}
\end{tabular}
\label{tab:event_examples} 
\end{table*}

\section{Dataset}
\label{sec:dataset}

\subsection{Video Collection}
\label{sec:video_collection}

The FIVR-200K dataset was designed with the following goals in mind: a) the videos should be associated with a large number of news events, b) the categories of these news events should be the same, and c) the dataset size needs to be sufficiently large to make retrieval of relevant results challenging. 

Based on the above requirements, we set up the process depicted in Figure \ref{fig:overview} to retrieve videos about major news events that took place during recent years. 
First, we crawled Wikipedia's `Current Event' page\footnote{\url{https://en.wikipedia.org/wiki/Portal:Current_events}} to build a collection of the major news events since the beginning of 2013. 
Each news event is associated with a topic, headline, text, date, and hyperlinks. 
Five examples of collected news events are displayed in Table \ref{tab:event_examples}. For the remaining steps of the process, we retained only news events categorized as `Armed conflicts and attacks' or `Disasters and accidents'. 
We selected these two categories to find multiple videos on YouTube that report on the same news event, and ultimately to collect numerous pairs of videos that are associated with each other through the relations of interest (DSV, CSV and ISV).  
The time interval used for crawling the news events was from January 1\textsuperscript{st} 2013 to December 31\textsuperscript{st} 2017. A total of 9,431 news events were collected, and 4,687 news events were retained after filtering.

In the next step, the public YouTube API\footnote{\url{https://developers.google.com/youtube/}} was used to collect videos by providing event headlines as queries. The results were filtered to contain only videos published at the corresponding event start date and up to one week after the event. 
Furthermore, they were filtered to contain only videos with a duration of up to five minutes, which resulted in the collection of 225,960 videos ($\sim$48 videos/event).
At this point, it is worth noting that several of the news event headlines in Wikipedia describe long-running news events (e.g., Syrian civil war), which is not an issue for our data collection process since the combination of the general event headline, and the particular event date is often sufficient to retrieve a variety of videos that depict the incidents of interest that are alluded by the respective Wikipedia entries.


\subsection{Query Selection}
\label{sec:query_selection}
Selecting ``appropriate'' queries is important for ensuring that the dataset annotations and evaluation protocol are representative of 
the challenges arising in real-world search tasks.
To this end, the query selection process was designed with two goals in mind: a) to generate challenging queries, i.e. queries that will lead to many distractor videos and challenge content-based retrieval systems, and b) to find queries that will lead to the retrieval of videos with various modifications that will not only be trivial NDV cases but also contain interesting variations (e.g., different viewpoints of the same scene), i.e., CSV and ISV. 
To achieve those two goals, we implemented a largely automatic process that is described below. 

First, the visual similarity between videos was computed as the cosine similarity between the tf-idf representations of their visual words.  
The visual words are derived from the NDVR method described in \cite{kordopatis2017b} and modified based on a Bag-of-Word (BoW) scheme. We sample one frame per second and extract the embedding vectors using a trained Deep Metric Learning (DML) network, which are then mapped and aggregated to the three closest visual words from a codebook of size 10k. 
The DML network was trained on the VCDB dataset\cite{jiang2014}, and the visual codebook was built by sampling one frame per video in the dataset and extracting the corresponding embedding vector.
Next, the textual similarity between videos was computed as the cosine similarity between the tf-idf representations of their titles. To perform the similarity calculation, we first pre-processed video titles with the NLTK toolkit \cite{bird2004}, applying part-of-speech (PoS) tagging, removing all verbs (which we found to introduce unnecessary noise) and providing the results to the NLTK WordNet-based lemmatizer to extract the lemmas, which constitute the word-based representation of the titles. 
The overall video similarity derives from the average of the visual and textual similarity. 
Bag of words was selected as a representation for both visual and text words because of its sparsity,  
which was practical for fast similarity calculation and efficient dataset annotation.

\begin{figure*}[t]
\centering
\includegraphics[width=15cm]{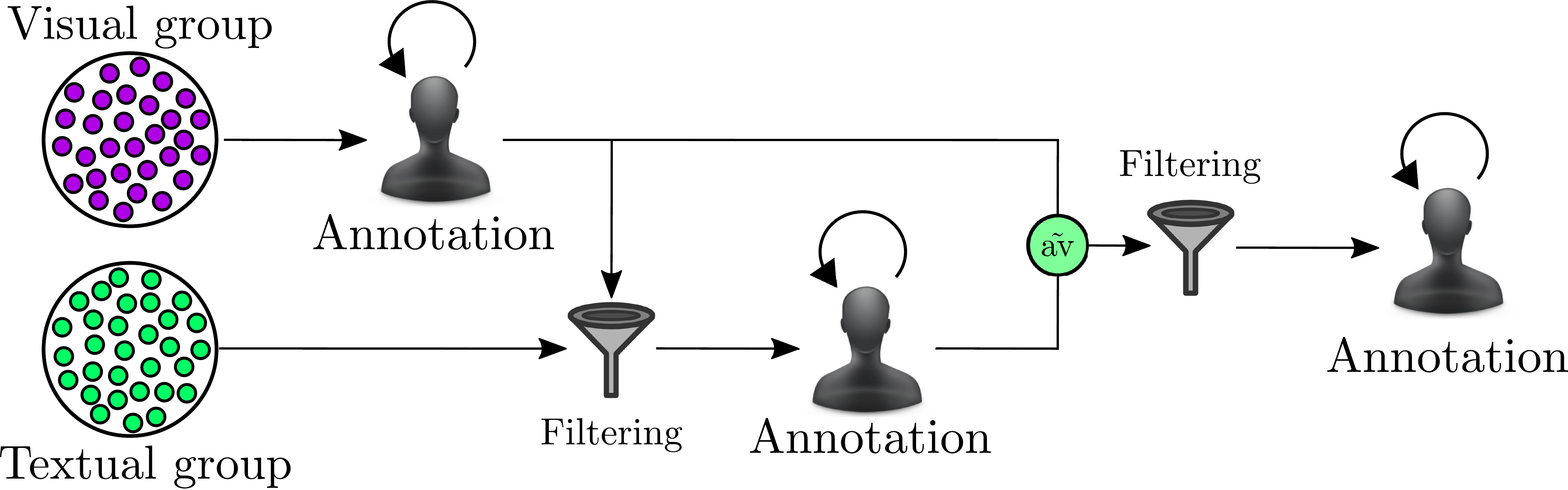}
\caption{Overview of the annotation process. Two groups of videos are created based on their visual and textual similarity to the query. Three annotation phases take place and two filtering steps are applied. $\tilde{\text{av}}$ stands for the average of visual and textual similarity between videos.}
\label{fig:annotation_overview}
\end{figure*}

In the next step, we computed all non-zero similarities between video pairs. 
Only video pairs that share at least one visual or text word were considered, which resulted in complexity much lower than $O(n^2)$. Afterwards, 
we created a video graph $G$ by connecting with an edge video pairs with similarity greater than a certain threshold $t_s$ (empirically set to 0.7). To identify meaningful video groups, we extracted the connected components $C$ of the video graph $G$ with more than two videos. Then, we defined the uploader ratio $r_c$ of each component $c\in C$ using Equation \ref{eq:uploader_ratio}.
\begin{equation}
	r_c = \frac{|\{u_v | v \in c, u_v \in U\}|}{N_c}
	\label{eq:uploader_ratio}   
\end{equation}
where the numerator is the number of unique uploaders in the component, $v$ is a video in the component, $u_v$ is the uploader of video $v$, $U$ is the set of uploaders in the dataset, and $N_c$ is the number of videos in the component. We empirically found that components with a low uploader ratio usually contain videos from a single specific channel (e.g., news channel) with titles that are very similar (e.g., exactly the same title with a different date) or with content that is visually highly similar (e.g., the same presenter reporting news in the same background). However, based on our definition, such videos are neither considered DSV nor CSV or ISV. For that reason, we discard components with an uploader ratio less than a threshold $t_r$ (empirically set to 0.75). Additionally, since we need components consisting of videos that refer to the same incident,  
we applied another criterion on the component set based on the publication date of their videos and retained only components consisting of videos that were published within a time window of two weeks.

Our goal was to find queries that lead to result sets with many DSV, CSV and ISV. Intuitively, large components with many (visually and textually) similar videos have a better chance of containing such videos. For that reason, we ranked connected components based on their size and selected one query video per component. We considered that short videos with few shots were the most suitable candidates for having been modified and reposted several times (both as single videos or as part of mash-ups). 
Therefore, we selected videos with a duration of less than a threshold $t_d$ (empirically set to 90 seconds). Attempting to find the original version of videos in each cluster, we 
chose the video that was published earliest as the query video.

The total number of queries using the above process was 635. Since it would be overly time consuming to annotate all of them, we selected the top 100 as the final query set (ranked based on the size of the corresponding graph component).

\subsection{Annotation Process}

Figure \ref{fig:annotation_overview} depicts the entire annotation process, which is carried out in three steps. Given a query, two groups of videos are retrieved, one based on visual similarity and one based on textual similarity. In the first step, we annotate the videos contained in the ``visual'' group. 
The end of the first step occurs when a total number of 100 irrelevant videos have been annotated after the last relevant result (i.e., annotated as NDV, DSV, CSV or ISV) or after the first 1000 videos have been annotated (whichever of the two criteria applies first). In the second step, videos in the ``textual'' group that have already been annotated as part of the visual group are removed. The annotation process continues with the remaining videos in the textual group. 
Similarly, this step ends either when a total number of 100 irrelevant videos have been annotated after the last relevant video or after the first 1000 videos have been annotated. To minimize the possibility of having missed relevant videos, in the third and final step, the remaining videos of the two groups are merged and filtered based on their publication date. We retained only videos that were published within a time window of a week before and after the publication date of the query.
These were ranked based on the average visual-textual similarity, and the annotation proceeded until either 100 irrelevant videos were found after the last relevant video, or no videos were left in the merged group. 

The annotation labels and corresponding definitions, which were used by the annotators, are as follows:
\begin{itemize}
\item\textbf{Near-Duplicate (ND)}: 
These are a special case of DSVs, as specified in Definition \ref{def:dsv}.

\item\textbf{Duplicate Scene (DS)}: DSVs are annotated with this label based on Definition \ref{def:dsv}.

\item\textbf{Complementary Scene (CS)}: CSVs are annotated with this label based on Definition \ref{def:csv}.

\item\textbf{Incident Scene (IS)}: ISVs are annotated with this label based on Definition \ref{def:isv}.

\item\textbf{Distractors (DI)}: Videos that do not fall in any of the above cases are annotated as distractors. 
\end{itemize}
For the annotation of the dataset, the extracted queries were split into two parts, each assigned to a different annotator. After the end of the annotation process, all annotated videos (excluding those labelled as DI) were revisited and tested for their consistency to the definitions by the lead author. 


\begin{figure}[t]
\centering
\subfigure[news events]{\includegraphics[width=8cm]{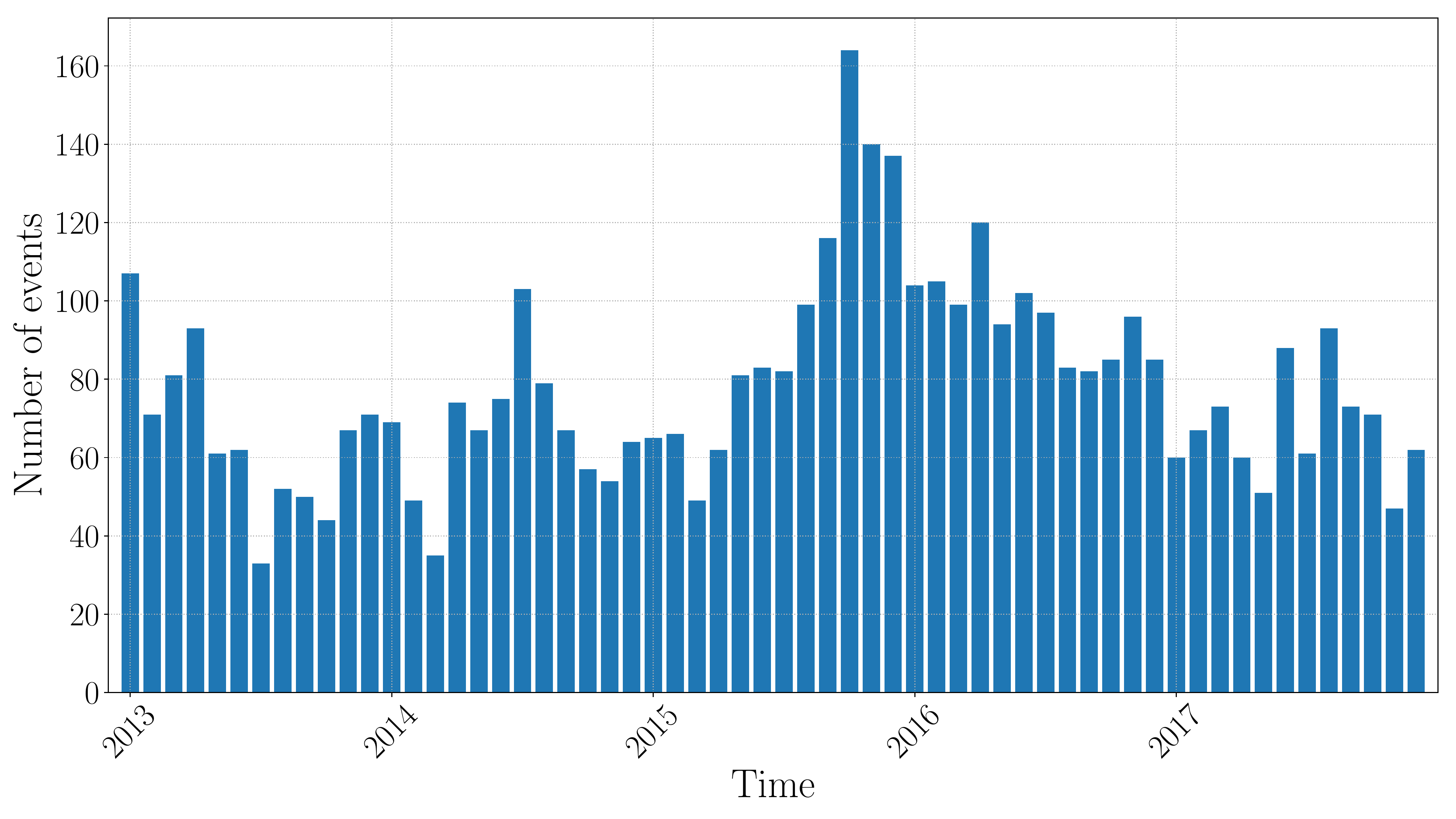}\label{fig:histograms_events}}
\subfigure[videos]{\includegraphics[width=8cm]{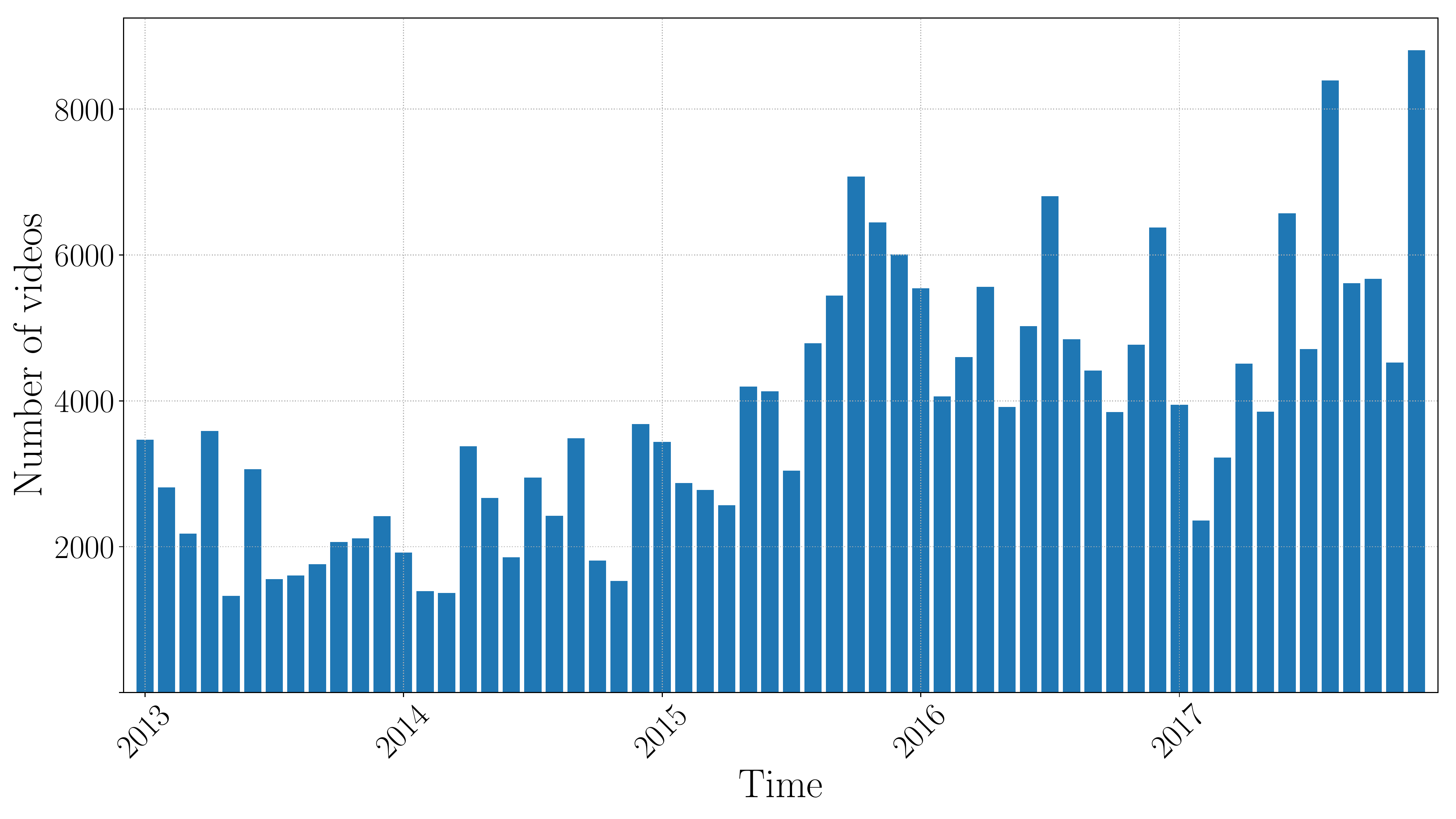}\label{fig:histograms_videos}}
\subfigure[queries]{\includegraphics[width=8cm]{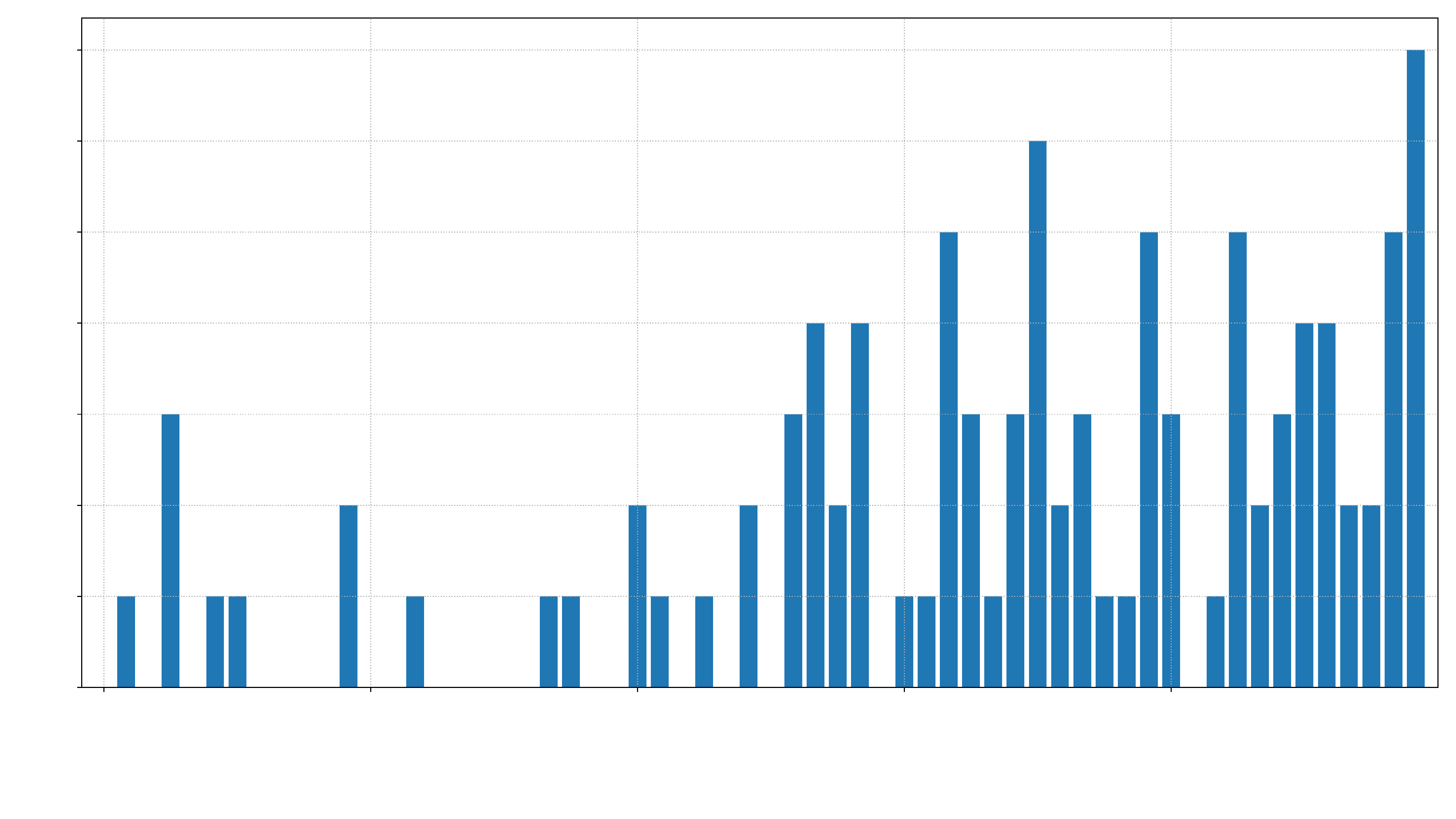}\label{fig:histograms_queries}}
\caption{Monthly distribution of a) news events, b) videos and c) queries.
}
\label{fig:histograms}
\end{figure}

\subsection{Dataset Statistics}
\label{sec:statistics}
In total, the dataset comprises 225,960 videos associated with 4,687 Wikipedia news events and 100 selected video queries. Figure \ref{fig:histograms} illustrates the monthly distribution of the collected news events, videos and queries. 
There is a noteworthy peak of news events during the last quarter of 2015. 
During that period, major wars (e.g., the Syrian civil war, the war in Afghanistan, the Yemeni civil war) and a number of devastating natural disasters (e.g., hurricane Joaquin, Hindu Kush earthquake and an intense Pacific typhoon season) took place leading to daily newsworthy incidents. From the temporal video distribution, one may notice an increase in video sharing in the last two years which does not correspond to the trend in the timeline of major news events.  
A possible explanation may be the increasing trend in video capturing and sharing on YouTube. 
Finally, 
it is noteworthy that the temporal distribution of queries approximately follows the one of videos over time with more query videos published during the last two years of the dataset. This confirms that the employed query selection process does not introduce temporal bias. 

\begin{figure}[t]
\centering
\includegraphics[width=8cm]{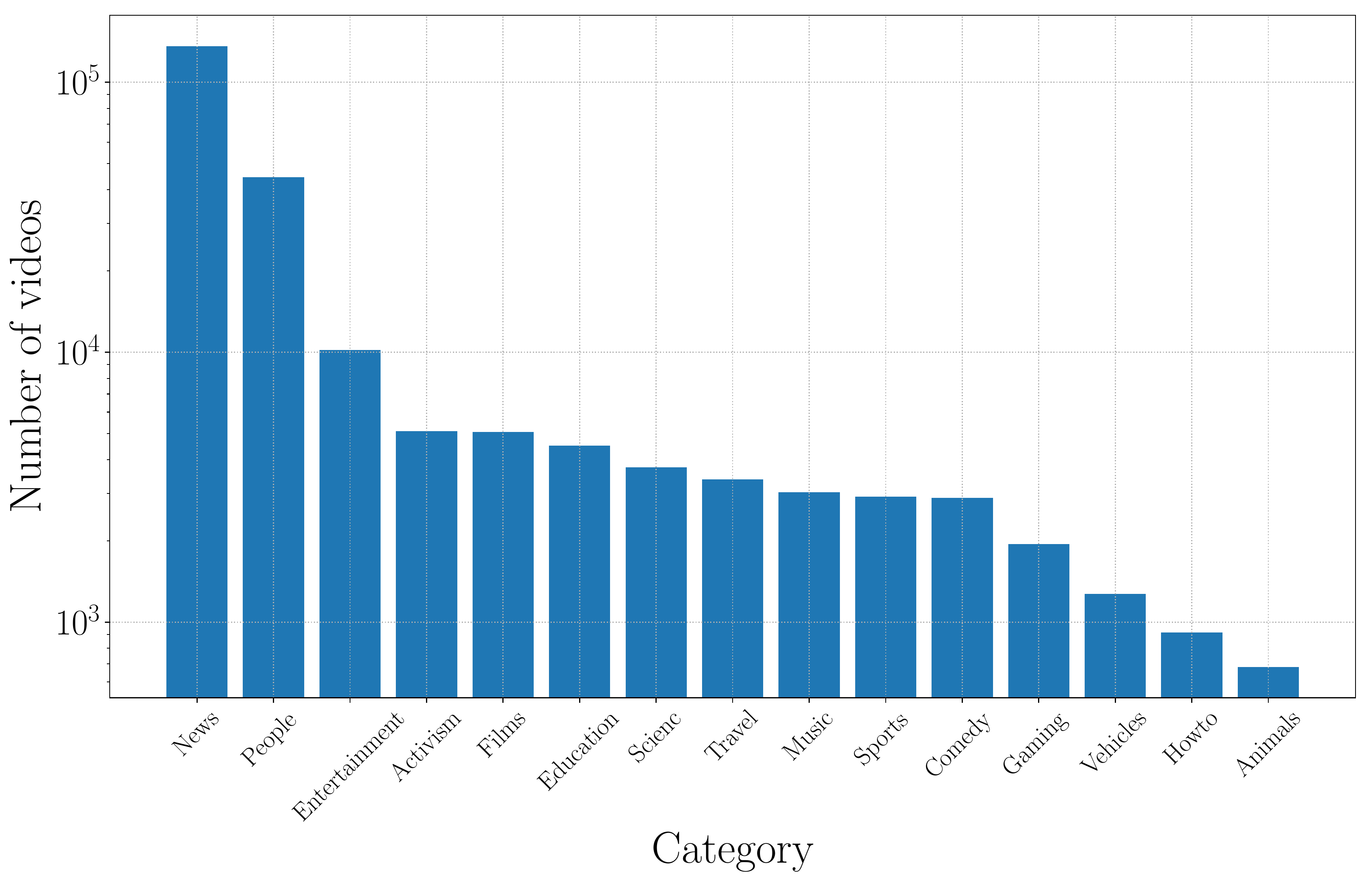}
\hspace{0.5cm}
\includegraphics[width=8cm]{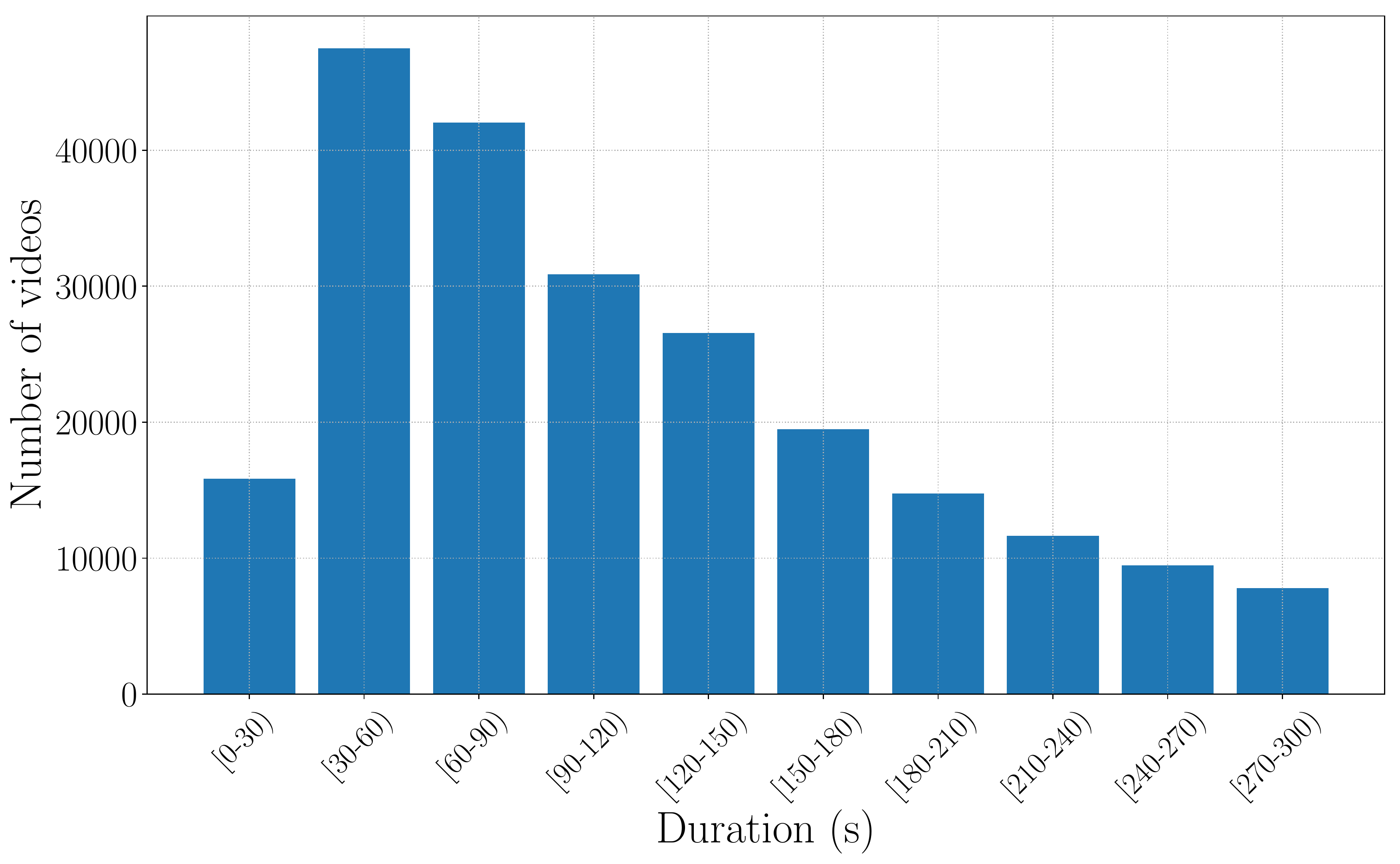}
\caption{Distribution of videos based on their category and duration.}
\label{fig:histogram_duration_categories}
\end{figure}

Table \ref{tab:top_event} presents the top news events based on their duration and number of collected videos. The duration of a news event is computed as the total number of days when it occurred in the collection. 
As expected, the longest news events, are wars or war-related events that usually last several years. 
The longest news event was the Syrian Civil War, which covered almost 500 days. However, news events with the most collected videos are breaking news events with large media coverage and live footage from multiple sources. The news event with the most collected videos was the terrorist attack in Paris, France on 13 November 2015, where multiple suicide bombers struck followed by several mass shootings. 
Figure \ref{fig:histogram_duration_categories} illustrates the distributions of video categories and duration. From the first, it is evident that the majority of collected videos are news related, which was expected due to the nature of the searched events. Additionally, the `People' category has a sizable portion of the collected videos. Regarding video duration, the majority of videos have a length between 30 to 120 seconds. 

\begin{table}[t]
  \caption {(left) the top 10 longest news events (right) the top 10 news events with the most videos.}
  \centering
  \scalebox{0.92}{
  \begin{tabular}{lc}
  \textbf{Long-running news events} & days \\ \hhline{==}
    Syrian civil war & 499\\
    War in Afghanistan & 250\\
    Iraqi insurgency & 137\\
    War in NW Pakistan & 118\\
    Iraqi civil war & 116\\
    War in Somalia & 101\\
    Yemeni civil war & 89\\
    Israel-Palestine conflict & 64\\
    War in Donbass & 62\\
    Libyan civil war & 61\\
  \end{tabular}
  \begin{tabular}{lc} 
  \textbf{Breaking news events} & videos \\ \hhline{==}
    November 2015 Paris attacks & 651\\
    2017 Atlantic hurricane season & 572\\
    Charlottesville riots & 569\\
    Charlie Hebdo shooting & 546\\
    2017 Las Vegas shooting & 542\\
    Umpqua College shooting & 486\\
    Assassination of Andrei Karlov & 476\\
    2016 central Italy earthquake & 475\\
    2014 Peshawar school massacre & 459\\
    2017 Manchester arena bombing & 457\\
  \end{tabular}
  }
  \label{tab:top_event}
\end{table}

To further delve into the dataset content, we processed the videos titles and extracted summary statistics. Initially, the language of the titles was detected using the detection approach by \cite{nakatani2010}. As expected, the predominant language was English with 81.16\%, followed by German with 2.58\%. It is noteworthy that Indonesian ranked third with 1.74\%, possibly due to several terrorist attacks that occurred in the region during the period of interest. Additionally, the most used nouns and locations are reported in Table \ref{tab:top_nouns_loc}. We extracted the nouns using the NLTK toolkit \cite{bird2004} and the mentioned countries using the method described in \cite{kordopatis2017c}. Unsurprisingly, the most used nouns were the ones related to wars and natural disasters, as well as the general words `news' and `video'. The most frequently mentioned countries were the ones related to long-lasting wars or major incidents with considerable media coverage. 

\begin{table}[h]
  \caption {(left) the top 10 most used nouns (right) the top 10 most refereed countries.}
  \centering
  \begin{tabular}{lc}
  \textbf{Nouns} & videos \\ \hhline{==}
    attack & 18192\\
    news & 12133\\
    earthquake & 8016\\
    fire & 7121\\
    hurricane & 6447\\
    crash & 6304\\
    video & 5790\\
    flood & 5394\\
    force & 4702\\
    army & 4464\\
  \end{tabular}
  \hspace{0.5cm}
  \begin{tabular}{lc} 
  \textbf{Locations} & videos \\ \hhline{==}
    Syria & 13952\\
    Ukraine & 4545\\
    Iraq & 4545\\
    Russia & 3990\\
    Yemen & 3988\\
    Turkey & 3653\\
    Israel & 2776\\
    Afghanistan & 2691\\
    China & 2604\\
    Egypt & 2306\\
  \end{tabular}
  \label{tab:top_nouns_loc}
\end{table}

In terms of content source, the dataset contains videos from 66,919 unique channels. As expected, the most prolific channels are news-related, including Wochit News, Ruptly, AP, and Al Jazeera, which constantly upload breaking-news content. 
Additionally, we grouped videos based on year of publication and found that the median of views per video remained approximately the same through the years. 

\begin{figure}[t]
\centering
\includegraphics[width=8cm]{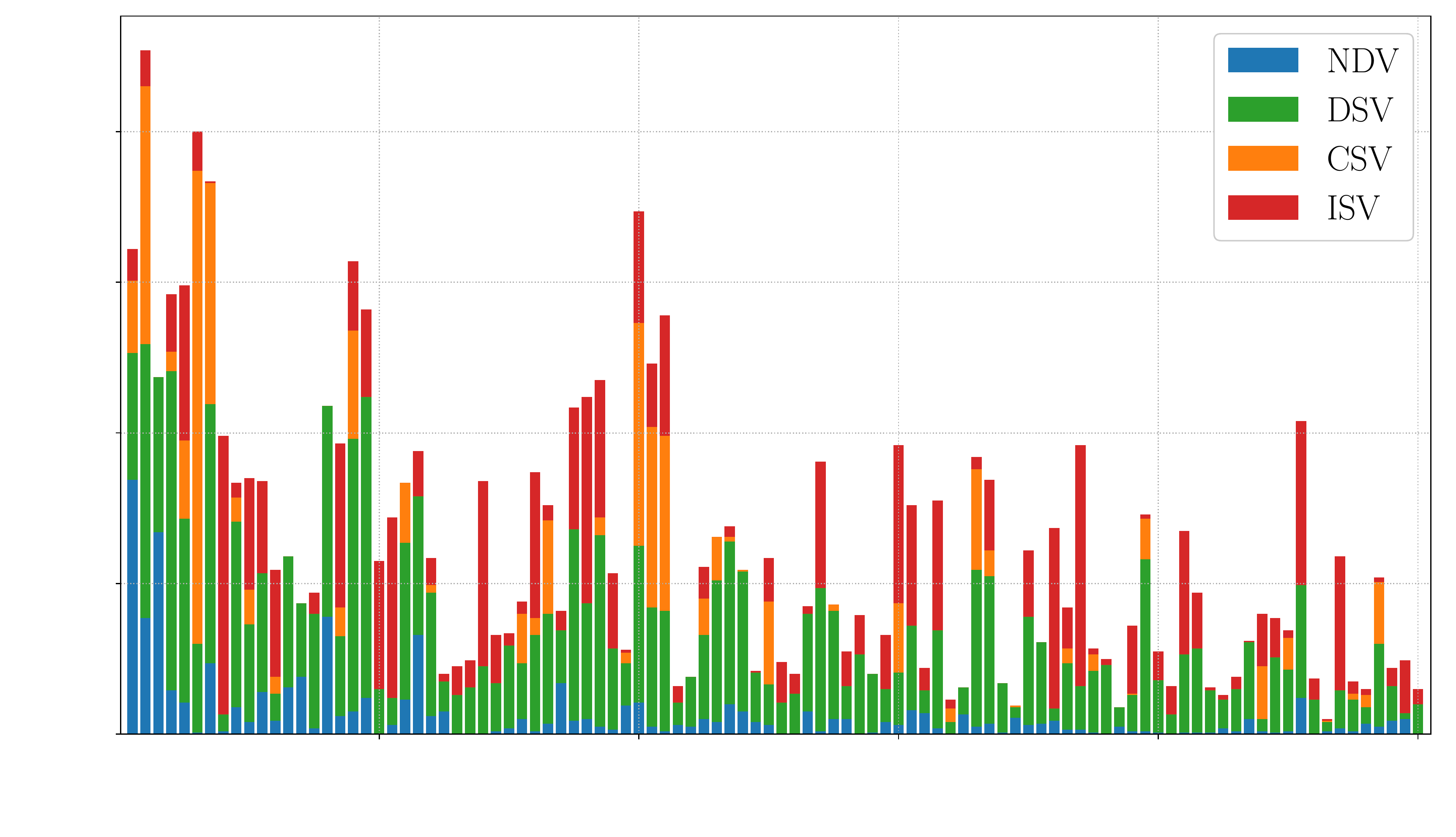}
\caption{Distribution of annotation labels per query (best viewed in colour).}
\label{fig:query_stacked}
\end{figure}

Regarding the annotation labels, we found that the selected queries have on average 13 NDV, 57 DSV, 18 CSV and 35 ISV. Figure \ref{fig:query_stacked} illustrates the distribution of annotation labels per query. 
Queries were ranked by the size of the cluster they were associated with (cf. Section \ref{sec:query_selection}). 
As expected, there was a considerable correlation (Pearson correlation=0.62) between cluster size and the number of videos that were annotated with one of the four relevant labels.
For all 100 queries, the total number of unique videos annotated (including DIs) was approximately 140 thousand. Some videos were annotated multiple times because they had different labels for different queries. 

\section{Comparative Study}
\label{sec:benchmark}

\subsection{Experimental Setup}
\label{sec:experimental_setup}

In this section, we conduct a comparative study to evaluate the performance of several state-of-the-art video retrieval systems. We compare five state-of-the-art approaches 
based on different feature extraction, aggregation and similarity calculation schemes of those presented in Section \ref{sec:soa_methods}. 
Additionally, three tasks are defined based on the labels that are considered relevant per task.

\subsubsection{Evaluation Metrics}
To evaluate retrieval performance, we build on the evaluation scheme described in \cite{wu2007}. We first employ the interpolated \textit{Precision-Recall} (PR) curve. 
\textit{Precision} is determined as the fraction of retrieved videos that are relevant to the query, while \textit{Recall} is the fraction of the total relevant videos retrieved 
(Equation \ref{eq:precision_recall}).
\begin{equation}
\begin{gathered}
Precision = \frac{TP}{TP+FP} \quad
Recall = \frac{TP}{TP+FN}
\end{gathered}
\label{eq:precision_recall}
\end{equation}
where $TP$, $FP$ and $FN$ are the true positives (correctly retrieved), false positives (incorrectly retrieved) and false negatives (missed matches), respectively. The interpolated PR-curve derives from averaging the Precision scores of all queries for given Recall ranges. The maximum Precision score is selected as the representative value for each Recall range. We further use \textit{mean Average Precision} (mAP) as defined in \cite{wu2007} to evaluate the quality of video rankings. For each query, the \textit{Average  Precision} (AP) is calculated based on Equation \ref{eq:map}.
\begin{equation}
AP = \frac{1}{n} \sum\limits_{i=0}^{n} \frac{i}{r_i}
\label{eq:map}
\end{equation}
where $n$ is the number of relevant videos to the query video, and $r_i$ is the rank of the $i$-th retrieved relevant video. The mAP is computed by averaging the AP scores across all queries.

\subsubsection{Benchmarked Approaches}

Of the feature aggregation and similarity calculation techniques described in Section \ref{sec:soa_methods}, we benchmark the following state-of-the-art approaches: 
\begin{itemize}
\item\textbf{Global Vectors}: 
In the approach by \cite{wu2007}, the HSV histograms are extracted for each video frame, and all frame descriptors are averaged to a single vector for the entire video. Video similarity is calculated based on the dot product between the respective vectors. This approach is denoted as GV.

\item\textbf{Bag-of-Words}: We select two methods using this feature aggregation scheme. The first \cite{cai2011} is a traditional BoW approach that employs the ACC\cite{huang1997} features as frame descriptors. Every frame descriptor is mapped to a single visual word of a visual codebook. The second approach \cite{kordopatis2017a} is a variant of the traditional BoW scheme based on the intermediate CNN features. The feature vectors extracted from each convolutional layer are mapped to a word of a visual codebook (one codebook per layer). For both methods, the final video representation is the \textit{tf-idf} representation of these visual words. Video ranking is performed based on the cosine similarity between the tf-idf video representations. The two methods are denoted as BoW and LBoW, respectively.

\item\textbf{Deep Metric Learning}: The approach by \cite{kordopatis2017b} is selected as representative of this feature aggregation scheme. The intermediate CNN features \cite{kordopatis2017a} are extracted from the video frames and combined into global video descriptors, similar to GV. These descriptors are fed to a DML network to calculate video embeddings. Video similarity is calculated based on the Euclidean distance between these embeddings. This approach is denoted as DML.

\item\textbf{Hashing Codes}: The approach by \cite{song2013} is selected as representative of this feature aggregation scheme. Multiple frame features are extracted, i.e., HSV and LBP \cite{zhao2007} and used to learn a group of hash functions that project video frames into the Hamming space and then combined to a single video representation. Hamming distance is employed to determine the similarity between videos. Here, we use the public implementation provided by the authors. This approach is denoted as HC.
\end{itemize}

For all methods, we extract one frame per second to generate the frame descriptors. For the Bag-of-Words methods, the codebooks are created by sampling one frame per video in the dataset and extracting their visual descriptors. The selection of appropriate codebook size is important, so we experimented with 1K and 10K visual words per codebook. Only the results for the 10K codebook size are presented, since there is a large performance gap in favour of 10K words. 
For the DML and HC, the most important tuning parameter is the dimensionality of the output vectors. Yet, from our experiments we concluded that it does not have a decisive impact on the performance of the approach. The DML is a supervised approach, so it is trained on the VCDB \cite{jiang2014} dataset. The HC is an unsupervised approach, but a sample of 50K frame descriptors is still required to learn a set of hash functions. An extensive evaluation of the sensitivity to the parameters of the benchmarked methods is beyond the scope of this work; hence we selected those parameter values suggested by the authors or ones that gave better results in our initial experiments.

\subsubsection{Visual Descriptors}

For a more comprehensive and fair comparison, we also implemented the benchmarked approaches with the following visual descriptors.

\begin{itemize}
\item\textbf{Hand-crafted Features}: We experiment with four widely used hand-crafted features in the literature: HSV histograms, LBP \cite{zhao2007}, ACC \cite{huang1997} and VLAD-SURF \cite{jegou2010}.

\item\textbf{Intermediate CNN Features}: We employ three popular architectures for the extraction of intermediate CNN features \cite{kordopatis2017a}: VGG-16 (VGG) \cite{simonyan2014}, ResNet-152 (RES) \cite{he2016} and Inception-V4 (INC) \cite{szegedy2017}.

\item\textbf{3D CNN Features}: We employ two popular architectures for the extraction of  3D CNN features: C3D \cite{tran2015} and I3D \cite{carreira2017}. To extract one visual descriptor per second, we feed the network with the corresponding number of frames suggested by the authors. We extract features with two techniques: (i) from the activations of the first fully connected layer after the convolutional layers, and (ii) from the intermediate 3D convolutional layers by applying MAC pooling in the spatial (similar to the CNN features) and temporal axis. 

\end{itemize}

The ResNet, Inception and I3D architectures are very deep, which made the utilization of all convolutional layers impractical. Hence, we extracted features from the activations of the convolutions before max-pooling. For the HC method, we set up three runs based on i) hand-crafted features, ii)  CNN features extracted from the three architectures, and iii) 3D CNN features extracted from the two architectures. For more information regarding the implementation of the benchmarked approaches, 
we direct readers to the Supplementary Material of the paper.

\subsubsection{Retrieval tasks}

We evaluate three different retrieval tasks. 
Table \ref{tab:acceted_labels} indicates the positive labels for each task. 
\begin{itemize}
\item\textbf{Duplicate Scene Video Retrieval (DSVR)}: this task represents the NDVR problem so it only accepts the videos annotated with ND or DS as relevant.

\item\textbf{Complementary Scene Video Retrieval (CSVR)}: this scenario is a strict variation of the FIVR problem where only the ND, DS and CS are accepted as relevant. 

\item\textbf{Incident Scene Video Retrieval (ISVR)}: this represents the general FIVR problem, and all labels (with the exception of DI) are considered relevant.
\end{itemize}

\begin{table}[bth]
  \centering
  \normalsize
  \caption {Positive labels for each evaluation setup.}
  \begin{tabular}{|c|} 
    \multicolumn{1}{c}{} \\ \hline
    \textbf{Task}        \\ \hline
    \textbf{DSVR}        \\ \hline
    \textbf{CSVR}        \\ \hline
    \textbf{ISVR}        \\ \hline
  \end{tabular}
  \begin{tabular}{|c|c|c|c|}
    \hline 
    \multicolumn{4}{|c|}{\textbf{Accepted Labels}} \\ \hline
    \textbf{ND} & \textbf{DS} & \textbf{CS} & \textbf{IS}    \\ \hline
    \checkmark  & \checkmark  & &  \\ \hline
    \checkmark  & \checkmark  & \checkmark &      \\ \hline
    \checkmark  & \checkmark  & \checkmark & \checkmark   \\ \hline
  \end{tabular}
  \label{tab:acceted_labels}
\end{table}

\subsection{Experiments}
\label{sec:experiments}

\subsubsection{Benchmarked approaches}

In this paragraph, we evaluate the performance of the five compared approaches. Table \ref{tab:app_map} illustrates the mAP of the benchmarked approaches on the three evaluation tasks of the FIVR-200K dataset and the CC\_WEB\_VIDEO dataset. LBoW outperforms all other approaches in all cases 
by a considerable margin. The second best performance is achieved by DML, followed by HC and BoW. GV had the worst results in all cases. 
In particular, LBoW achieves a mAP score of 0.710 in the DSVR task, followed by DML and HC with 0.398 and 0.265 respectively. BoW and GV are the two worst performing approaches with 0.240 and 0.165 mAP values, respectively. For the CSVR task, all approaches exhibit a drop in mAP, between 0.018 and 0.04. The performance is significantly worse in the ISVR task for all benchmarked approaches. The best method (LBoW) achieves a mAP score of 0.572, whereas the worst (GV) only 0.118.

\begin{table}[b]
  \centering
  \caption {mAP of the benchmarked approaches for the three retrieval tasks and the CC\_WEB\_VIDEO dataset.}
  \begin{tabular}{|c|c|c|c?c|}
    \hline
    \textbf{Run}       & \textbf{DSVR} & \textbf{CSVR}  &  \textbf{ISVR} &  \textbf{CC\_WEB} \\ \hhline{=====}
    \textbf{GV}\cite{wu2007}               & 0.165  & 0.153  & 0.118  & 0.892  \\ \hline
    \textbf{BoW}\cite{cai2011}             & 0.240  & 0.220  & 0.171  & 0.944  \\ \hline
    \textbf{LBoW}\cite{kordopatis2017a}   & \textbf{0.710}  & \textbf{0.675}  & \textbf{0.572} & \textbf{0.976}  \\ \hline
    \textbf{DML}\cite{kordopatis2017b}     & 0.398  & 0.378  & 0.309  & 0.971  \\ \hline
    \textbf{HC}\cite{song2013}            & 0.265  & 0.247  & 0.193  & 0.958  \\ \hline
  \end{tabular}
\label{tab:app_map}
\end{table}

The results make clear that the DSVR task of the proposed framework is closely related to the NDVR problem which is simulated by the CC\_WEB\_VIDEO dataset. It is evident that the performances of all methods on FIVR-200K are significantly lower compared to CC\_WEB\_VIDEO, highlighting that the newly proposed dataset is much more challenging. All methods report very high mAP scores on CC\_WEB\_VIDEO, achieving values as high as 0.976. Even the GV approach achieves a score close to 0.9. 
The main reason for the performance gap is that the vast majority of positive video pairs in FIVR-200K are partially similar, not in their entirety but in particular segments. Additionally, FIVR-200K contains a wide variety of user-generated videos about news events of similar nature resulting to many challenging distractors. 

\subsubsection{Comprehensive experiments}

\begin{table}[t]
  \caption {mAP of the benchmarked approaches and the different visual features for three retrieval tasks.}
  \centering
  \scalebox{1}{
  \begin{tabular}{|c|c|c|c|c|c|}
    \hline
    \multicolumn{6}{|c|}{\textbf{DSVR}} \\ \hline
    \textbf{Run}  & \textbf{GV} & \textbf{BoW}  &  \textbf{LBoW} & \textbf{DML} & \textbf{HC}  \\ \hhline{======}
    \textbf{HSV}          & 0.165  & 0.202   & N/A    & 0.163  & \multirow{4}{*}{0.360}   \\ \cline{1-5}
    \textbf{LBP}         & 0.112  & 0.158   & N/A    & 0.097  &   \\ \cline{1-5}
    \textbf{ACC}         & 0.196  & 0.240   & N/A    & 0.182  &   \\ \cline{1-5}
    \textbf{VLAD}        & 0.294  & 0.323   & N/A    & 0.285  &   \\ \hhline{======}
    \textbf{VGG}         & 0.366  & 0.575   & \textbf{0.710}  & 0.398 & \multirow{3}{*}{0.470}   \\ \cline{1-5}
    \textbf{RES}         & 0.350  & 0.523   & 0.596  & 0.374 &    \\ \cline{1-5}
    \textbf{INC}         & 0.333  & 0.500   & 0.608  & 0.367 &    \\ \hhline{======}
    \textbf{C3D$_{fc}$}  & 0.244  & 0.341   & N/A    & 0.266 &    \multirow{4}{*}{0.434}   \\ \cline{1-5}
    \textbf{C3D$_{int}$} & 0.355  & 0.541   & 0.658  & 0.387 &    \\ \cline{1-5}
    \textbf{I3D$_{fc}$}  & 0.321  & 0.464   & N/A    & 0.336 &    \\ \cline{1-5}
    \textbf{I3D$_{int}$} & 0.366  & 0.574   & 0.665  & 0.425 &    \\ 
    \hline
  \end{tabular}}
  \scalebox{1}{
  \begin{tabular}{|c|c|c|c|c|c|c|}
    \multicolumn{6}{c}{} \\ \hline
    \multicolumn{6}{|c|}{\textbf{CSVR}} \\ \hline
    \textbf{Run}  & \textbf{GV} & \textbf{BoW}  &  \textbf{LBoW} & \textbf{DML} & \textbf{HC}  \\ \hhline{======}
    \textbf{HSV}          & 0.153  & 0.189   & N/A    & 0.150  & \multirow{4}{*}{0.339}   \\ \cline{1-5}
    \textbf{LBP}         & 0.106  & 0.146   & N/A    & 0.091  &   \\ \cline{1-5}
    \textbf{ACC}         & 0.183  & 0.220   & N/A    & 0.169  &   \\ \cline{1-5}
    \textbf{VLAD}        & 0.275  & 0.311   & N/A    & 0.265  &   \\ \hhline{======}
    \textbf{VGG}         & 0.347  & 0.543   & \textbf{0.675}  & 0.378 & \multirow{3}{*}{0.454}   \\ \cline{1-5}
    \textbf{RES}         & 0.333  & 0.499   & 0.572  & 0.358  &    \\ \cline{1-5}
    \textbf{INC}         & 0.313  & 0.473   & 0.571  & 0.348  &    \\ \hhline{======}
    \textbf{C3D$_{fc}$}  & 0.231  & 0.314   & N/A    & 0.252  & \multirow{4}{*}{0.415} \\ \cline{1-5}
    \textbf{C3D$_{int}$} & 0.336  & 0.502   & 0.628  & 0.374  &    \\ \cline{1-5}
    \textbf{I3D$_{fc}$}  & 0.312  & 0.444   & N/A    & 0.325  &    \\ \cline{1-5}
    \textbf{I3D$_{int}$} & 0.345  & 0.544   & 0.634  & 0.405  &    \\
    \hline
  \end{tabular}}
  \scalebox{1}{
  \begin{tabular}{|c|c|c|c|c|c|c|}
    \multicolumn{6}{c}{} \\ \hline
    \multicolumn{6}{|c|}{\textbf{ISVR}} \\ \hline
    \textbf{Run}  & \textbf{GV} & \textbf{BoW}  &  \textbf{LBoW} & \textbf{DML} & \textbf{HC}  \\ \hhline{======}
    \textbf{HSV}          & 0.118  & 0.143   & N/A    & 0.116  & \multirow{4}{*}{0.262}   \\ \cline{1-5}
    \textbf{LBP}         & 0.087  & 0.113   & N/A    & 0.074  &    \\ \cline{1-5}
    \textbf{ACC}         & 0.142  & 0.171   & N/A    & 0.128  &    \\ \cline{1-5}
    \textbf{VLAD}        & 0.214  & 0.236   & N/A    & 0.206  &    \\ \hhline{======}
    \textbf{VGG}         & 0.281  & 0.450   & \textbf{0.572}  & 0.309 & \multirow{3}{*}{0.382}   \\ \cline{1-5}
    \textbf{RES}         & 0.274  & 0.414   & 0.488  & 0.296  &    \\ \cline{1-5}
    \textbf{INC}         & 0.257  & 0.406   & 0.488  & 0.290  &    \\ \hhline{======}
    \textbf{C3D$_{fc}$}  & 0.176  & 0.242   & N/A    & 0.194  & \multirow{4}{*}{0.334}   \\ \cline{1-5}
    \textbf{C3D$_{int}$} & 0.261  & 0.398   & 0.510  & 0.295  &    \\ \cline{1-5}
    \textbf{I3D$_{fc}$}  & 0.253  & 0.364   & N/A    & 0.265  &    \\ \cline{1-5}
    \textbf{I3D$_{int}$} & 0.280  & 0.450   & 0.527  & 0.332  &    \\ 
    \hline
  \end{tabular}}
  \label{tab:map_fivr}
\end{table}

Table \ref{tab:map_fivr} presents the mAP performance of all possible feature-aggregation combinations.

To begin with the DSVR task, similar to the previous section, the LBoW aggregation scheme in combination with the VGG CNN features achieves the best result (mAP$=0.710$) at a considerable margin from the second. Notably, VGG performs consistently better than the other two CNN architectures for all aggregation schemes. Additionally, LBoW clearly outperforms the regular BoW aggregation irrespective of CNN or 3D CNN architecture. The same conclusions apply in the case of 3D CNN features. The intermediate I3D features achieve the best results for all methods, with performance close to or better than the performance of VGG features. For instance, in the case of DML, the I3D$_{int}$ achieves 0.425 mAP, while VGG 0.398. Among the handcrafted features, VLAD-SURF provides the best results (mAP$=0.323$); however, the performance gap with deep features is considerable.

Similar conclusions apply in the case of the CSVR task, with the LBoW-VGG combination achieving the best results (mAP$=0.675$). The performance for all runs decreases slightly compared to the DSVR task, indicating that it presents a more challenging problem. 

The performance is notably worse in the case of the ISVR task for every approach-feature combination, with the decrease ranging from 0.03 to 0.13 in mAP. This reveals that ISVR is a much more challenging problem and new systems need to be devised to effectively address it. Overall, deep network features (either CNN or 3D CNN) outperform the handcrafted features by a significant margin. Moreover, DML boosts the performance of deep features compared to the GV runs. However, this is not the case for handcrafted features where the performance drops. 
Moreover, for 3D CNN architectures, the runs with intermediate features consistently report better performance compared to the runs with features from the fully connected layers. HC in combination with CNN features achieves the best performance compared to the other feature bundles, for all evaluation tasks. Additionally, GV performs poorly for all features compared to the other three schemes. For the rest of this paper, we are going to refer to each method in relation to its combination with the best-performing features, i.e., VGG features for GV, BoW and LBoW, I3D$_{int}$ features for DML, and the CNN features for HC.

\begin{figure}[t]
\centering
\includegraphics[width=6cm]{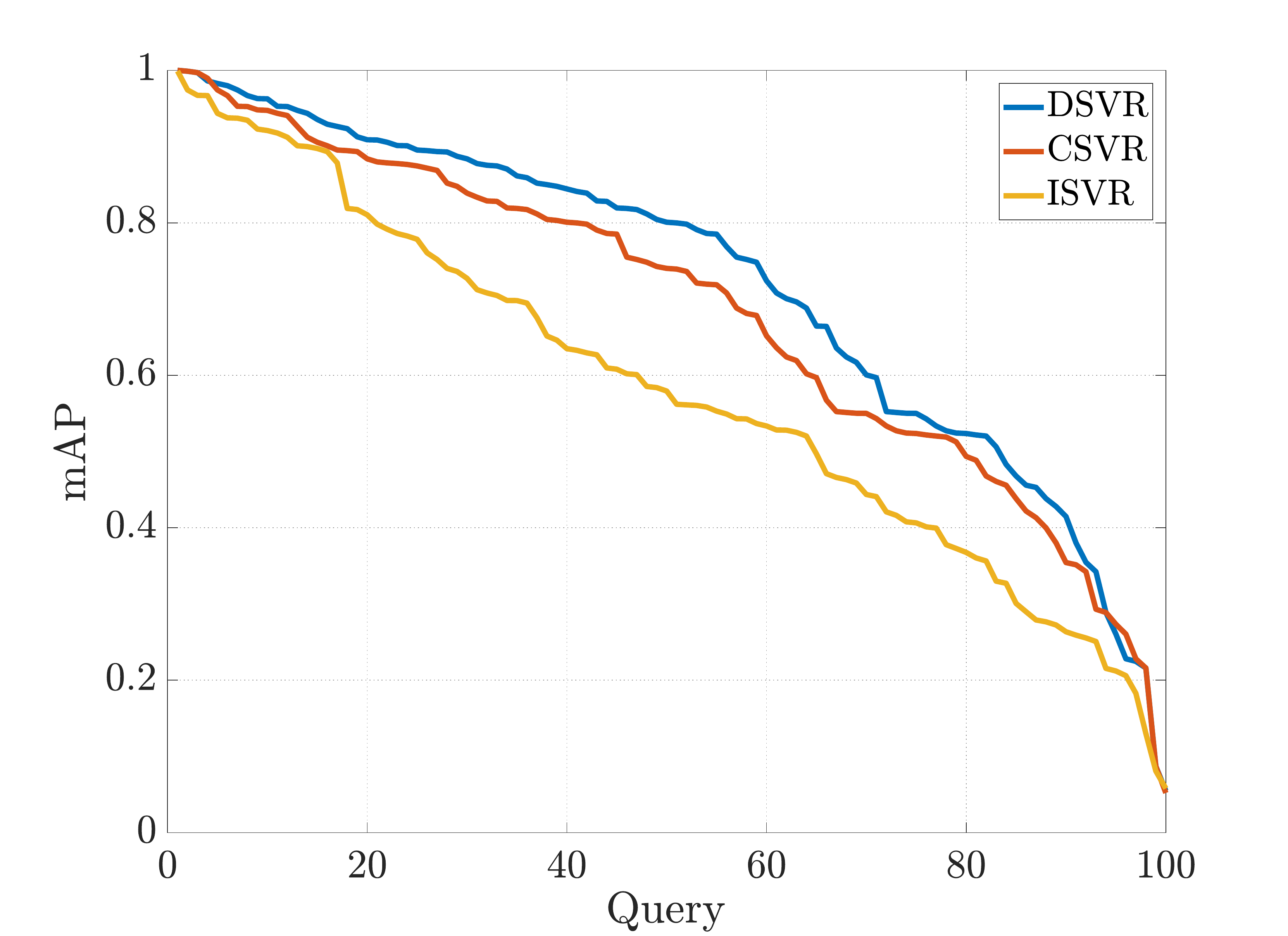}
\caption{mAP of the queries in the dataset based on LBoW with VGG features run for the three retrieval tasks. The queries are ranked in descending order.}
\label{fig:query_map}
\end{figure}

\begin{figure*}[t]
\centering
\small
\subfigure[DSVR]{\includegraphics[width=5.5cm]{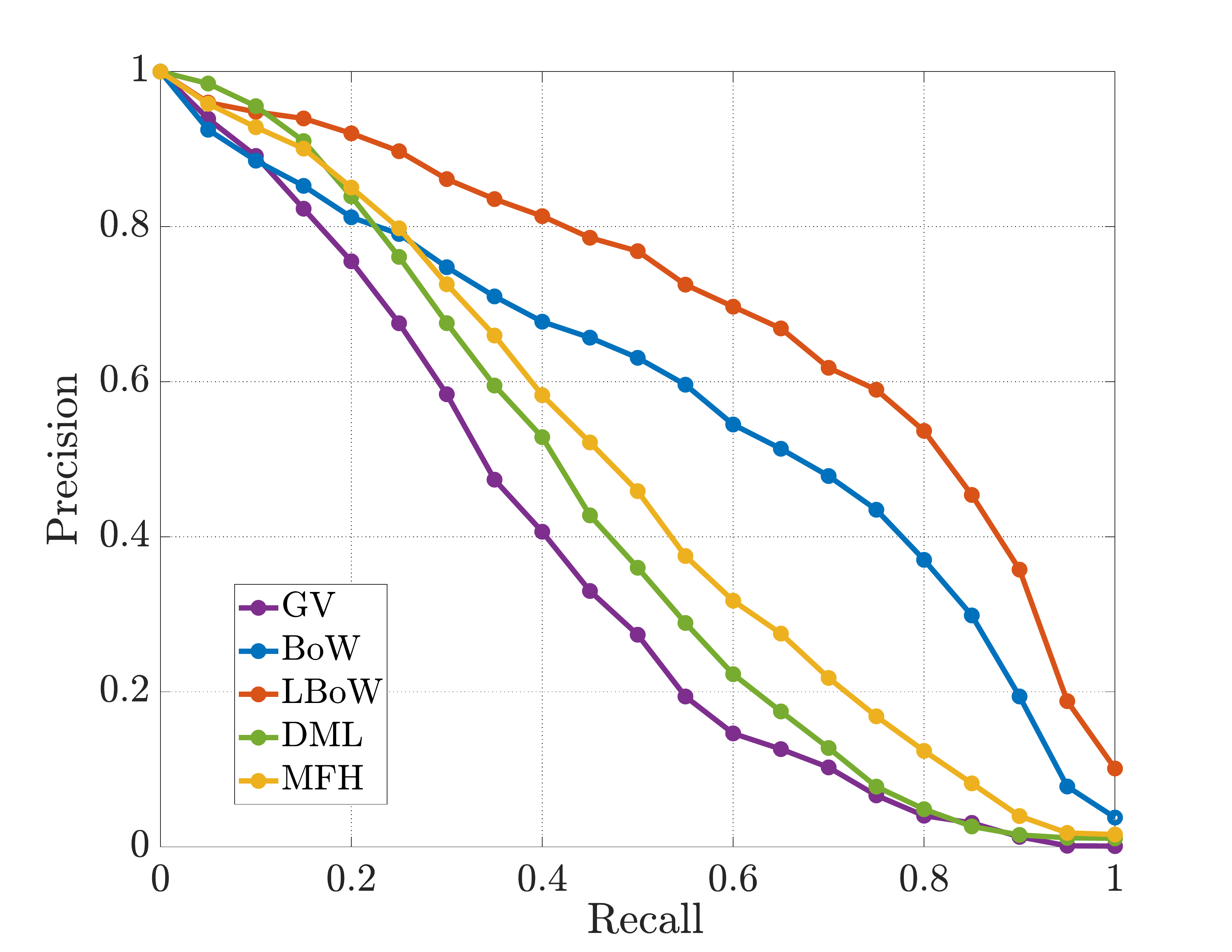}}
\subfigure[CSVR]{\includegraphics[width=5.5cm]{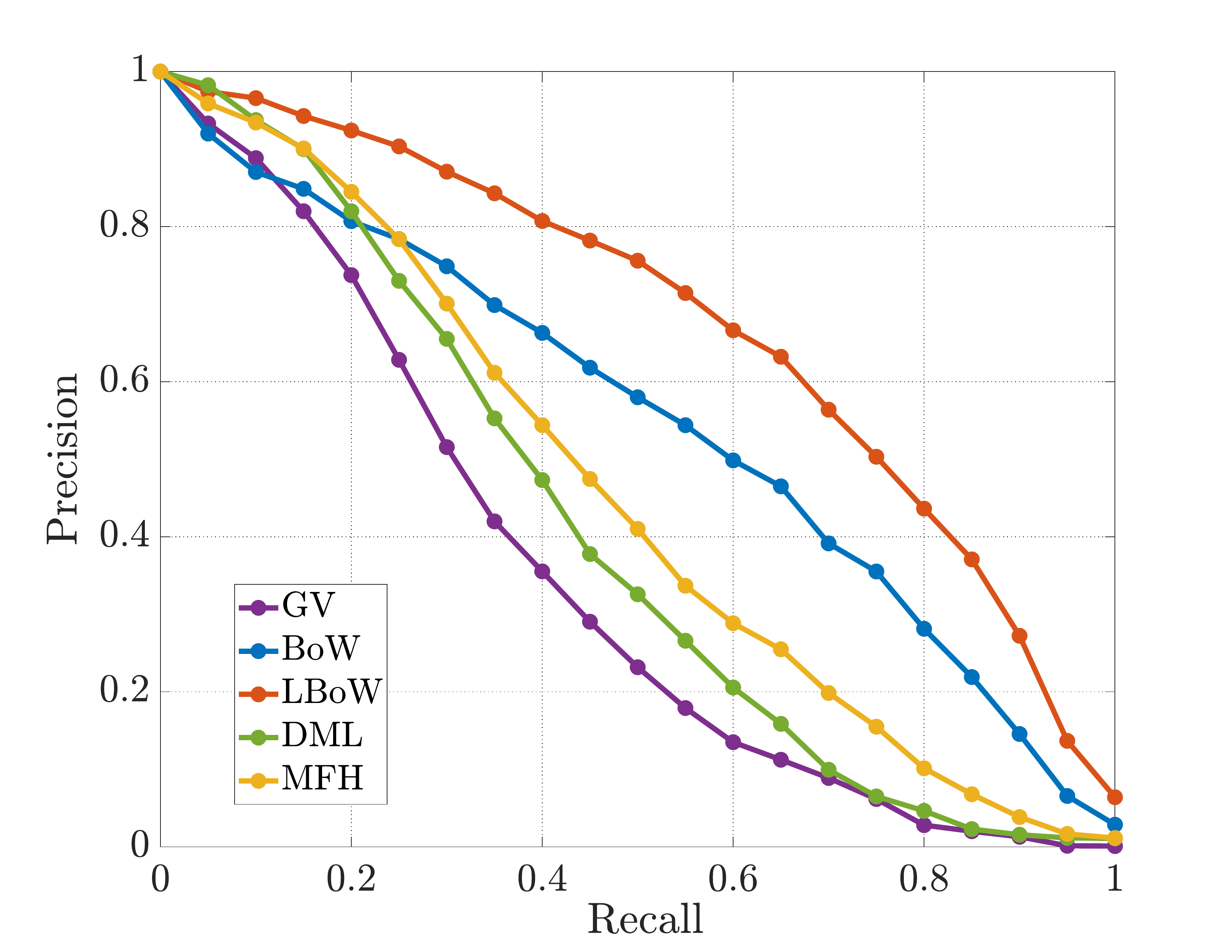}}
\subfigure[ISVR]{\includegraphics[width=5.5cm]{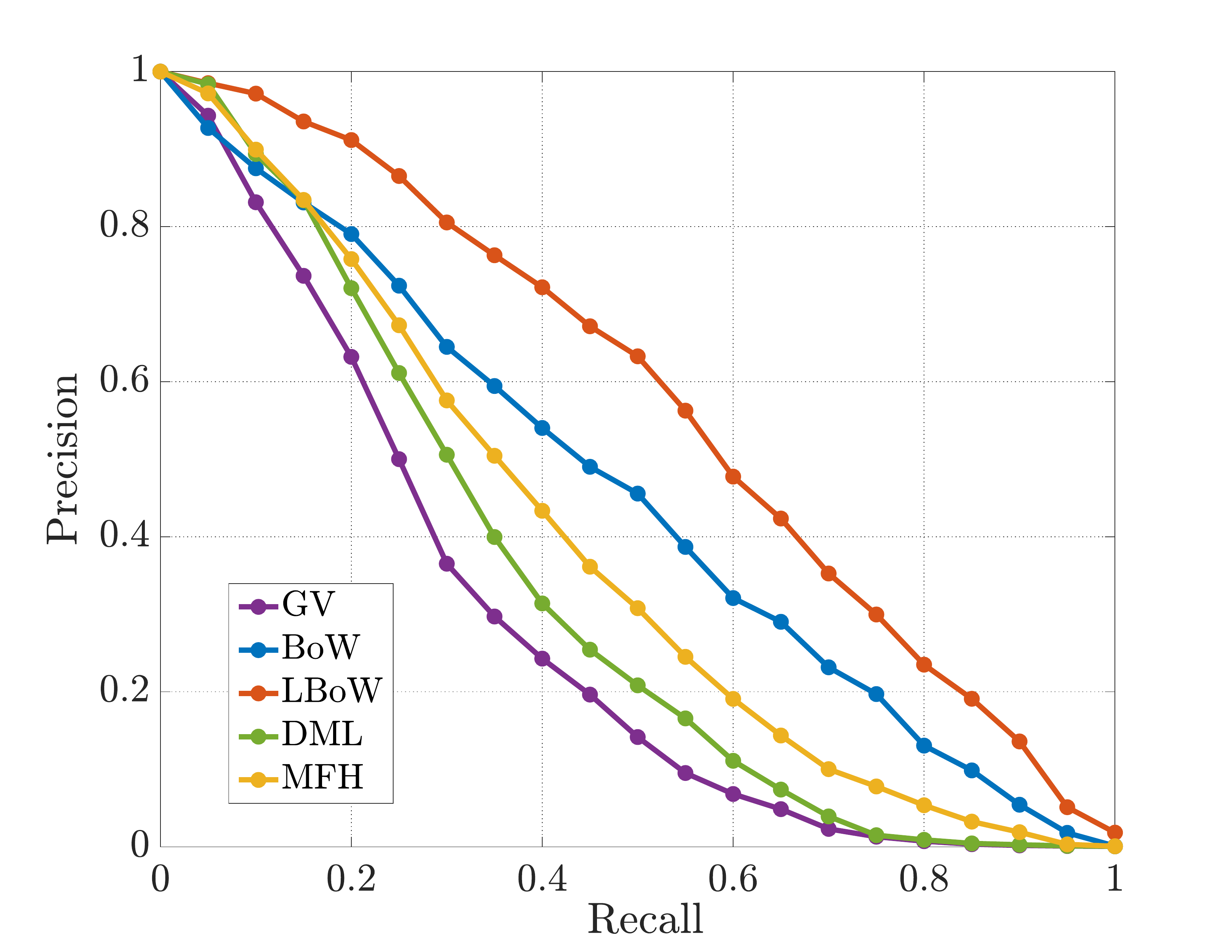}}
\caption{Interpolated PR-curves of the best-performing features for each approach in the three retrieval tasks.}
\label{fig:pr_curves}
\end{figure*}

Figure \ref{fig:query_map} illustrates the mAP per query of the best-performing run (LBoW with VGG features) for the three different tasks. The queries are ranked in descending order based on their mAP. For the DSVR task, 50\% of the queries achieve higher than a 0.8 mAP, while the performance is significantly lower for the remaining queries. There is a notable drop in performance in the CSVR task, with 80\% of the queries having higher than a 0.5 mAP. Finally, it is evident that ISVR is a much harder task than the other two, with the majority of queries having lower than 0.6 mAP.

Figure \ref{fig:pr_curves} illustrates the interpolated PR-curves of the best-performing runs for each method and for each evaluation task. Similar conclusions apply as in the case of their mAP comparison. 
LBoW outperforms other runs consistently for all three tasks by a significant margin. However, there is a large gap between BoW and the other three runs. A reasonable explanation is that the BoW representation retains local information from the video frames, in contrast to the other aggregation methods that average frame descriptors in a global video representation. This is of critical importance for all three tasks since only a minority of similar videos share their entire content to the queries. Similar to the mAP evaluation, GV performs poorly for all retrieval tasks compared to the other schemes.

Table \ref{tab:method_times} presents the requirements in terms of storage space and computation time for the best-performing run of each method. The results of all methods have been measured using the open source library Scikit-learn \cite{van2014} in Python, on a Linux PC with a 4-core i7-4770K and 32GB of RAM. It is noteworthy that LBoW's superior performance comes at a high computational and storage cost. In particular, it needs approximately 1.2s per query to perform retrieval (being the slowest among the five approaches) and 3KB per video to store the video representations. The fastest method is the HC with 51ms per query, followed by BoW with three times slower retrieval time. The method that requires the least memory space in RAM is BoW reserving only 209B per video. DML is in the middle of the rank for both measures. The most demanding method in terms of storage space is GV requiring approximately 16KB for each video descriptor. Note that these figures are derived from computing video similarities for one query at a time, without vectorizing all query descriptors in a single matrix. This practice would significantly decrease retrieval time for all methods. 

\begin{table}[t]
  \centering
  \caption {Storage and computation requirements per video for the best-performing run for each approach. The storage requirements are measured in bytes (B) and the retrieval time in milliseconds (ms).}
  \begin{tabular}{|l|c|c|c|c|c|}
    \hline
    \textbf{Method}       & \textbf{GV} & \textbf{BoW}  &  \textbf{LBoW} &  \textbf{DML} & \textbf{HC} \\ \hline
    \textbf{Storage space}        & 16,384 & 209    & 3,050 & 2,048 & 512    \\ \hline
    \textbf{Retrieval time}       & 499    & 152     & 1,155   & 333   & 51  \\ 
    \hline
  \end{tabular}
\label{tab:method_times}
\end{table}


\subsubsection{Within-dataset retrieval}

Our initial goal for the construction of the FIVR-200K is to be used for evaluation purposes in its entirety. 
However, it is not always possible to have access to a separate dataset that simulates the same or a similar retrieval problem. To overcome this issue, we have also devised a within-dataset experimental setup, where we split the dataset into two separate video sets, one for the development/training of the methods and one for evaluation\footnote{The dataset split is only applied in this subsection.}. To do so, we order the videos based on their publication time and then split them in half, resulting in two sets of videos from different time periods. 
We select the early period video set for training and the late period video set for testing. The total number of queries in the training and test set are 31 and 69 respectively. 

\begin{table}[h]
  \centering
  \caption {mAP of the benchmarked approaches built based on the FIVR-200K training set and evaluated on the FIVR-200K test set for the three retrieval tasks.}
  \begin{tabular}{|c|c|c|c|}
    \hline
    \textbf{Run}       & \textbf{DSVR} & \textbf{CSVR}  &  \textbf{ISVR} \\ \hhline{====}
    \textbf{GV}        & 0.389  & 0.370   & 0.301 \\ \hline
    \textbf{BoW}       & 0.302  & 0.287   & 0.237 \\ \hline
    \textbf{LBoW}      & 0.362  & 0.344   & 0.280 \\ \hline
    \textbf{DML}       & 0.465  & 0.443   & 0.381 \\ \hline
    \textbf{HC}       & \textbf{0.468}  & \textbf{0.444}   & \textbf{0.382} \\ \hline
  \end{tabular}
\label{tab:split_exp}
\end{table}

Table \ref{tab:split_exp} presents the performance of the benchmarked approaches in the three evaluation tasks. There is a considerable decrease in terms of mAP for BoW and LBoW runs reaching approximately half their performance compared to the previous runs for all three tasks ($\approx$46-51\%). We observed similar decreases in performance when using VCDB for development (i.e., generation of visual codebooks) and the whole FIVR-200K for testing. This makes clear that BoW-based schemes are quite sensitive to the dataset that is used for generating the underlying visual codebooks. 
There is also a negligible drop in performance for the HC scheme (less than 0.01 in terms of mAP); hence, in this setting HC achieves the best results among all methods for all three tasks. As expected, DML is boosted when using part of the FIVR-200K for training. The improvement for DML in all tasks ranges between 0.03 and 0.05. 
Finally, the GV approach also sees a small improvement in all evaluation tasks compared to the initial results. 

\section{Conclusion}
\label{sec:conclusion}

In this paper, we introduced the problem of Fine-grained Incident Video Retrieval (FIVR). First, we provided definitions for the various types of video associations arising in the more general problem setting of FIVR. 
Next, we built a large-scale dataset, FIVR-200K, with the aim of addressing the benchmarking needs of the problem. The dataset comprises 225,960 YouTube videos, 
collected based on approximately 5,000 global news events crawled from Wikipedia over five years (2013-2017). Then, we selected 100 queries based on a principled approach that automatically assessed the suitability of a query video for performing evaluations for the current problem. 
We also devised a protocol for annotating the dataset according to four labels for video pairs. Finally, we conducted a thorough experimental study on the dataset comparing five state-of-the-art methods, six feature extraction methods and five CNN/3D CNN architectures and four video descriptor aggregation schemes. For the benchmark, we considered three retrieval tasks that represented different instances of the problem and accepted different labels as relevant, i.e., DSVR, CSVR and ISVR. The best-performing methods achieved mAP scores of 0.710, 0.675 and 0.572, respectively. 

In general, retrieval performance across all experiments was not very high compared to performance values that have been reported for related datasets, such as CC\_WEB\_VIDEO. This demonstrates that the proposed problem and associated dataset offer a challenging setting with considerable room for improvement, especially in the case of the ISVR task. 

We see several opportunities for future research on improved methods. Global video representations appear to have strong limits in terms of performance; one way to improve performance is to investigate frame-level matching practices that take into account the temporal alignment between videos. 
Additionally, end-to-end training approaches of CNN or 3D CNN networks could have a significant impact on the performance of systems, especially in the case of the ISVR task. Finally, query expansion approaches have been successfully applied on various retrieval problems, so they could potentially offer an option for achieving further improvements. However, all suggested alternatives are computationally complex, so they have to be carefully designed to be practical in a large-scale dataset such as FIVR-200K.

Due to its large size and wide variety of user-generated videos and news events, FIVR-200K could also facilitate many similar research problems, such as event reconstruction and synchronization. In the future, we will consider extending the dataset annotation to cover the needs of such problems. 

\section{Acknowledgements}

This work has been supported by the InVID and WeVerify projects, partially funded by the European Commission under contract numbers 687786 and 825297, respectively.

\balance

\bibliographystyle{IEEEtranS}
\bibliography{ref}

\end{document}